\documentclass[journal,twocolumn,twoside]{IEEEtranTCOM}
\usepackage{graphicx}
\usepackage{epsfig}
\usepackage{epstopdf}
\usepackage{amsmath,amsthm,amsfonts,amssymb}
\usepackage{multirow}
\usepackage{cite}
\usepackage{verbatim}
\usepackage{tikz}
\usetikzlibrary{arrows}
\usetikzlibrary{backgrounds}
\usepackage{ellipsis}
\usepackage{xcolor}
\usepackage{xifthen}
\usepackage{algorithm}
\usepackage{algpseudocode}
\usepackage{cancel}
\usepackage{float}
\usepackage{subfig}

\theoremstyle{definition} 
\newtheorem{prop}{Proposition}
\theoremstyle{definition} 
\theoremstyle{definition} 

\usetikzlibrary{shapes,arrows,intersections}
\usetikzlibrary{patterns,decorations}
\usetikzlibrary{mindmap}
\usetikzlibrary{calc}
\usepackage{ellipsis}
\usetikzlibrary{decorations.pathreplacing,decorations.markings,shapes.geometric,arrows,decorations.pathmorphing,decorations.footprints,fadings,calc,trees,mindmap,shadows,decorations.text,patterns,positioning,shapes,matrix,fit,chains}
\usepackage{arydshln}

\tikzset
{
	BlocksStyle/.style =
	{
		shape			= rectangle,			% shape
		rounded corners	= 0.0cm,				% radius of the rounded corner
		minimum height	= 0.7cm,				% | minimum size of the node
		minimum width	= 0.9cm,				% |
		rotate			= 0,					% angle of rotation
		scale			= 1.0,					% scaling factor
		draw			= black,				% draw the border with this color
		line width		= 0.00cm,				% thickness
		align			= center,				% text alignment
		text			= black,				% color of the fonts
		font			= \normalsize\normalfont,	% shape and dimension of the font
		inner xsep		= 0.2cm,				% min. dist. btw text and borders along x dimension
		inner ysep		= 0.2cm,				% min. dist. btw text and borders along x dimension
	}
}

\tikzset
{
	BlocksStyleb/.style =
	{
		shape			= rectangle,			% shape
		rounded corners	= 0.0cm,				% radius of the rounded corner
		minimum height	= 0.7cm,				% | minimum size of the node
		minimum width	= 0.9cm,				% |
		rotate			= 0,					% angle of rotation
		scale			= 1.0,					% scaling factor
		draw			= black,				% draw the border with this color
		line width		= 0.02cm,				% thickness
		align			= center,				% text alignment
		text			= black,				% color of the fonts
		font			= \normalsize\normalfont,	% shape and dimension of the font
		inner xsep		= 0.2cm,				% min. dist. btw text and borders along x dimension
		inner ysep		= 0.2cm,				% min. dist. btw text and borders along x dimension
	}
}

\tikzset
{
	WideBlocksStyle/.style =
	{
		BlocksStyle,
		text width		= 2.0cm,				% max. width of the text
	}
}
\pgfmathsetseed{1}
\usetikzlibrary{shapes,positioning,arrows,calc,graphs}
\usetikzlibrary{decorations.pathreplacing,decorations.markings,shapes.geometric}
\tikzset{naming/.style={align=center,font=\small}}
\tikzset{antenna/.style={insert path={-- coordinate (ant#1) ++(0,0.25) -- +(135:0.25) + (0,0) -- +(45:0.25)}}}
\tikzset{station/.style={naming,draw,shape=dart,shape border rotate=90, minimum width=10mm, minimum height=10mm,outer sep=0pt,inner sep=3pt}}
\tikzset{mobile/.style={naming,draw,shape=rectangle,minimum width=12mm,minimum height=6mm, outer sep=0pt,inner sep=3pt}}
\tikzset{radiation/.style={{decorate,decoration={expanding waves,angle=90,segment length=4pt}}}}

\tikzset
{
	HighlightingStyle/.style =
	{
		color				= blue,	
		line width			= 0.04cm,			% thickness
		arrows				= -,				% starting-ending arrows
		dotted,
	}
}	% no semicolons are here required

\tikzset
{
	HighlightingStyleD/.style =
	{
		color				= blue,	% color
		line width			= 0.04cm,			% thickness
		arrows				= -,				% starting-ending arrows
		dashed,
	}
}	% no semicolons are here required

\tikzset
{
	HighlightingStyleB/.style =
	{
		color				= black,	% color
		line width			= 0.04cm,			% thickness
		arrows				= -,				% starting-ending arrows
		solid,
	}
}

\tikzset
{
	HighlightingStyleE/.style =
	{
		color				= red,	% color
		line width			= 0.05cm,			% thickness
		arrows				= -,				% starting-ending arrows
		dashed,
	}
}	
\tikzset
{
	HighlightingStyleC/.style =
	{
		color				= black,
		fill=white,
		text=black,	% color
		line width			= 0.05cm,			% thickness
		arrows				= -,				% starting-ending arrows
 		rounded corners		= 0.0cm,			%
		solid,
	}
}

\tikzset
{
	LinesStyle/.style =
	{
		color				= black,	% color
		line width			= 0.02cm,			% thickness
		solid,
	}
}	

\tikzset
{
	LinesStyleC/.style =
	{
		color				= black,	% color
		line width			= 0.08cm,			% thickness
		solid,
	}
}

\tikzset
{
	LinesStyleR/.style =
	{
		color				= black,	% color
		line width			= 0.08cm,			% thickness
		densely dotted,
	}
}

\tikzset
{
	LinesStyleE/.style =
	{
		color				= red,	% color
		line width			= 0.06cm,			% thickness
		solid,
	}
}

\tikzset
{
	LinesStyleb/.style =
	{
		color				= black,	% color
		line width			= 0.02cm,			% thickness
		arrows				= -latex',			% starting-ending arrows
 		shorten <			= 0.1cm,			% shorten the ending point
		solid,
	}
}

\tikzset
{
	SumNodesStyle/.style =
	{
		shape			= circle,				% shape
		minimum size	= 0.1cm,				% | minimum size of the node
		rotate			= 0,					% angle of rotation
		scale			= 0.6,					% scaling factor
		draw			= black,				% draw the border with this color
		line width		= 0.02cm,				% thickness
		text			= black,				% color of the fonts
		font			= \normalsize\normalfont,	% shape and dimension of the font
	}
}

\DeclareMathOperator{\F2}{\mathbb{F}_2}
\DeclareMathOperator{\sgn}{sgn}
\DeclareMathOperator{\sort}{sort}

\newcommand{\bplus}{\mathrel{\sum\!\!\!\!\boxplus}}

\newcommand{\Iset}{\mathcal{I}}

\begin{document}

\title{Multi-Kernel Polar Codes:\\ Concept and Design Principles}

\author{\IEEEauthorblockN{ Valerio Bioglio, Fr\'ed\'eric Gabry, Ingmar Land, Jean-Claude Belfiore\\}
\IEEEauthorblockA{Mathematical and Algorithmic Sciences Lab\\ France Research Center, Huawei Technologies France SASU\\
Email: $\{$valerio.bioglio,frederic.gabry,ingmar.land,jean.claude.belfiore$\}$@huawei.com}} 

\maketitle

\begin{abstract}
In this paper, we propose a new polar code construction by employing kernels of different sizes in the Kronecker product of the transformation matrix, thus generalizing the original construction by Arikan. 
The proposed multi-kernel polar code allows for more flexibility in terms of the code length, moreover allowing for various new design principles.  
We describe in detail encoding as well as successive cancellation (SC) decoding and SC list (SCL) decoding, and we provide a novel design method for the frozen set that allows to optimise the performance under list decoding, as opposed to original relability-based code design.  
Finally, we numerically demonstrate the advantage of multi-kernel polar codes under the new design principles compared to punctured and shortened polar codes. 
\end{abstract}

\section{Introduction}
\label{sec:intro}
Polar codes are a family of error-correcting codes recently introduced by Arikan in \cite{polar} as the first codes able to provably achieve channel capacity for a large number of channels. 
The construction of a polar code of length $N=2^n$ is based on the recursive concatenation of the binary matrix $T_2 = \bigl(\begin{smallmatrix} 1 & 0 \\ 1 & 1 \end{smallmatrix} \bigr)$, referred to as the kernel of the transformation.  
This operation results in a transformation matrix $T_N = T_2^{\otimes n}$, given by the $n$-fold Kronecker power of the kernel matrix $T_2$, converting the physical channel into $N$ virtual synthetic channels characterized by either very high or very low reliability.  
This channel polarization effect leads to a portion of fully reliable channels that tends to the channel capacity for symmetric binary-input memoryless channels, when the code length tends to infinity.
In the asymptotic case, successive cancellation (SC) decoding is sufficient to achieve the channel capacity \cite{polar}.  
In the finite-length regime, successive cancellation list (SCL) decoding \cite{list_decoding} leads to performance competitive with many other classes of channel codes, like LDPC codes, particularly when an outer CRC code or generalizations thereof are applied \cite{CRC_aided}.  
Due to their excellent performance, polar codes were recently adopted as moderate-length codes for 5G \cite{polar_5G}.

\begin{figure}[t]
    \center
	\includegraphics[width=0.42\textwidth]{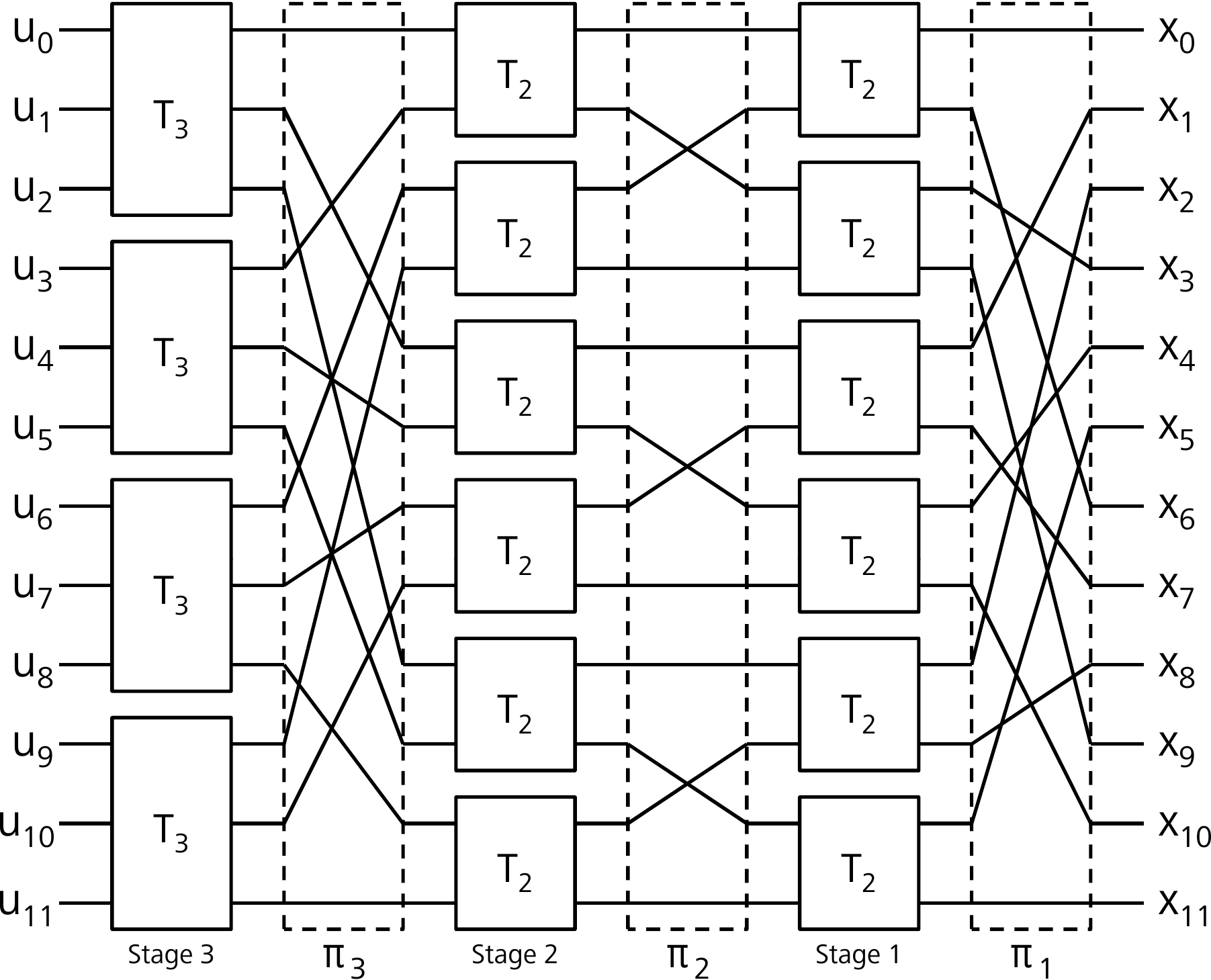}
	\caption{Tanner graph of the multi-kernel polar code of length $N=12$ for the transformation matrix $T_{12} = T_2 \otimes T_2 \otimes T_3$.}
	\label{fig:G_12}
\end{figure}

As conjectured in \cite{polar}, the polarization phenomenon obtained by the Kronecker powers of $T_2$ can be extended to other kernels. 
In \cite{exp_urbanke}, authors show necessary and sufficient conditions for kernels to polarize, allowing researchers to propose larger binary kernels as bases of novel polar codes \cite{kernel_presman}. 
Non-binary kernels have been proposed to improve the asymptotic error probability \cite{non_bin_ker,nb_polar}, while mixing binary and non-binary kernels showed additional improvement over homogeneous kernels constructions \cite{mixed_kernel}. 
As a result, current constructions restrict the length of polar codes to be in the form $N = l^n$. 

This code length constraint can be a huge limitation to practical use of polar codes, since only few block lengths can be expressed as a power of an integer. 
Punctured \cite{chen_kai_punc} and shortened \cite{wang_liu} polar codes have been proposed to increase the number of achievable block lengths. 
Even if these techniques offer a practical way to construct codes of arbitrary lengths, they show many disadvantages. 
In fact, punctured and shortened codes are decoded by means of their mother polar codes, increasing the decoding latency with respect to the actual code length. 
Moreover, the location of dummy bits, altering the polarization of the codes, has to be carefully chosen to avoid catastrophic error-rate performance \cite{isit_punc}. 
Finally, the lack of structure between the frozen sets and the puncturing or shortening patterns complicates the code design \cite{punct_paper}.

In this paper, we present multi-kernel polar codes, which generalize polar codes by mixing kernels of different sizes over the same binary alphabet. 
These codes were theorized in \cite{exp_pol_kron}, where their polarization rate is calculated algebrically, and explicitaly constructed in \cite{mk_arxiv}; they  conceptually permit to construct polar codes of any block length while keeping the polarization effect \cite{MK_pol_proof}. 
The encoding follows the general structure of polar codes, and the decoding can be performed by successive cancellation as well. 
Building on our previous contributions in \cite{mk_arxiv} and \cite{mk_dist_arxiv}, in this paper we present a thorough description and analysis of construction of multi-kernel polar codes, discussing the issues related to the choice of component kernels.  
Finally, similarly to \cite{polar_to_RM} we combine the aforementioned designs into a novel hybrid design exhibiting good error correction performance at moderate length. 

This paper is organized as follows. 
In Section \ref{sec:model}, we present the general construction, including the encoding and decoding, of multi-kernel polar codes. 
In Section \ref{sec:kernel} we provide recommendations for the selection of kernels to be used in the multi-kernel construction. 
In Section \ref{sec:design} we describe explicitly the different designs for multi-kernel polar codes, namely by reliability, by minimum distance, and by a hybrid criterion.
In Section \ref{sec:examples} we discuss the performance of the proposed codes, and Section \ref{sec:conclusions} concludes this paper.

\section{Code and Decoder}
\label{sec:model}
In this section, we introduce the structure, encoding and decoding of multi-kernel polar codes.  As an example, the Tanner graph of a code of length $N=12$ is depicted in Fig.~\ref{fig:G_12}, comprising two kernels of size $2$ and one kernel of size $3$.  

%--------------------------------------------------------
\subsection{Code Structure and Encoding}
%--------------------------------------------------------

Multi-kernel polar codes are a generalization of Arikan's polar codes obtained by using binary kernels of different sizes in the construction of the transformation matrix of the code. 
An $(N,K)$ multi-kernel polar code of length $N$ and dimension $K$ is defined by a $N \times N$ transformation matrix 
\begin{equation}
  T_N = T_{p_1} \otimes T_{p_2} \otimes \dots \otimes T_{p_s} ,
  \label{equ:GN-MK}
\end{equation}
with $N = p_1 \cdot p_2 \cdot \dots \cdot p_s$, and a frozen set $\mathcal{F} \subset [N]$, where $[N] = \{0,1,2,\ldots,N-1\}$, such that $|\mathcal{F}| = N-K$.  The information set is defined as $\mathcal{I} = \mathcal{F}^C$.
Building blocks of the code are the $p_i \times p_i$ matrices $T_{p_i}$ with binary entries, which define kernels of dimension $p_i$ \cite{MT_qary}. 
A list of binary polarizing kernels with maximum exponents, i.e. maximum polarization, can be found in \cite{non_bin_ker}.  
However, other kernels may be advantageous for the design of multi-kernel polar codes, as described in Sec.~\ref{sec:kernel}.
The frozen set collects the indices of the input vector to be frozen: its design will be discussed in Section~\ref{sec:design}. 
Codewords $x \in \F2^N$ are generated from the input vector $u \in \F2^N$ by $x = u \cdot T_N$, where $u_j = 0$ for $j \in \mathcal{F}$ and the remaining $K$ entries of $u_i$ for $i \in \mathcal{I}$ store the information to be transmitted. 
Note that outer CRCs or similar parity checks may be inserted as for original polar codes. 
In the following, we will refer to polar codes when the transformation matrix is generated using a single kernel according to the original formulation by Arikan, while we will refer to multi-kernel polar codes if more than one binary kernel is used in the transformation matrix generation. 

The order of the kernels in \eqref{equ:GN-MK} is important for the design of the code, as this operation is not commutative. 
Changing the order of kernels in $T_N$ is equivalent to permuting its rows and columns, since for any Kronecker product there exist two permutation matrices $P,Q$ such that $A \otimes B = P \cdot ( B \otimes A ) \cdot Q$ \cite{math-matrix-book}.
In practice, changing the order of the kernels leads to a transformation matrix in which rows and columns are permuted as compared to the original transformation matrix. 
Every frozen set imposed on the original matrix can hence be mapped in a frozen set of the permuted matrix: all the kernel orders lead to equivalent codes. 
However, this order may have an effect on the polarization of the virtual channels as discussed in Section~\ref{sec:reliability}.

%--------------------------------------------------------
\subsection{Tanner Graph}
%--------------------------------------------------------

The structure of multi-kernel polar codes can be illustrated by the Tanner graph as depicted in Figure~\ref{fig:G_12}. 
This graph describes the transformation matrix $T_N$ of the code, and consists of various $p_i \times p_i$ boxes, each corresponding to a kernel $T_{p_i}$ which defines the relation between the input vector and the output vector. 
A $p_i \times p_i$ box has $p_i$ inputs and outputs. 
A stage of the graph corresponds to a factor in the Kronecker product of the transformation matrix, and is depicted by $N/p_i$ boxes vertically distributed, for a total of $s$ stages, counted from the codeword $x$ (right) to the input vector $u$ (left).  

The connections between stages are implicitly defined by the Kronecker product. 
Stage $i$ has $N/p_i$ boxes, each one representing a kernel $T_{p_i}$, that are connected to the $N/p_{i-1}$ boxes of stage $i-1$ through an edge permutation $\pi_i$. 
These permutations operate in blocks, where two boxes in different blocks are not connected. 
Denoting $N_i = \prod_{j=1}^{i-1} p_j$ the partial product of the kernel sizes up to stage $i$, with $N_1=1$, we can divide the boxes forming stages $i$ and $i-1$ in $N/N_{i+1}$ blocks. 
Inside a block, boxes of stages $i-1$ are further divided into $N_i$ sub-blocks, so that the $j$-th box of stage $i$ is connected to the $j$-th output of each sub-block of stage $i-1$. 
This canonical permutation $\rho_i$, depicted in Equation \eqref{eq:Q_can}, will be used as a basis to create the general permutations between stages. 
In fact, given the canonical permutation $\rho_i$, the permutation $\pi_i$ is given by $\pi_i = \left( \rho_i | \: \rho_i+N_{i+1} | \: \rho_i+2N_{i+1} | \: \dots | \: \rho_i+(N/N_{i+1}-1)N_{i+1} \right)$ for $i = 2,\dots,s$. 
Note that for the last stage, we have $\pi_s = \rho_s$. 
First permutation $\pi_1$, acting like the bit-reversal permutation for polar codes, is an exception obtained inverting the product of the other permutations as $\pi_1 = (\pi_2 \cdot \dots \cdot \pi_s)^{-1}$.

\begin{figure*}[!t]
	\normalsize
	\begin{equation}
		\label{eq:Q_can} 
		\setcounter{MaxMatrixCols}{14}
		\rho_i = 
		\begin{pmatrix}
		    1 & 2     & \dots & N_i          & N_i+1 & N_i+2 & \dots & (n_i-1)N_i+1 & \dots & N_{i+1} \\[0.5ex]
		    1 & p_i+1 & \dots & (N_i-1)p_i+1 & 2     & p_i+2 & \dots & p_i          & \dots & N_{i+1} 
		\end{pmatrix}
	\end{equation}
	\hrulefill
	\vspace*{4pt}
\end{figure*}

%--------------------------------------------------------
\subsection{Decoding}
%--------------------------------------------------------

Decoding of multi-kernel polar codes is performed by successive cancellation (SC) \cite{polar} on the Tanner graph of the code. 
Similar to polar codes,  enhanced SC-based decoding methods, like simplified SC (SSC) \cite{SSC}, SC list (SCL) \cite{list_decoding} or SC stack decoding \cite{pc_stack} may be employed. 
In SC, bits are decoded sequentially using the log-likelihood ratios (LLRs) of the received symbols along with the (estimated) previously decoded bits. 
If $\lambda_i$ corresponds to the LLR of input bit $u_i$, an SC decoder sequentially evaluates 
\begin{equation}
\label{SC_eq}
\lambda_i = f_i^N( l_0 , l_1 , \dots , l_{N-1} , \hat{u}_0 , \hat{u}_1 , \ldots, \hat{u}_{i-1} )
\end{equation}
for every $i$ from $0$ to $N-1$, where $l_i$ corresponds to the LLR of the codebit $x_i$. 
In the following, we assume a BPSK transmission over an AWGN channel, referred to as the BI-AWGN, denoting with $E_s$ the energy of the transmitted symbols and with $N_0$ the single-sided noise power density. 
With constellation points $\pm 1$, the SNR may be given in $E_s/N_0$ or in $E_b/N_0 = 1/R \cdot E_s/N_0$, where $R$ denotes the code rate and $\sigma^2 = N_0/(2 E_s)$ is the variance of the AWGN.  
With $y = [y_0,\dots,y_{N-1}]$ denoting the output of the BI-AWGN channel, the channel LLRs are computed as $l_i = 2y_i/\sigma^2$. 

The recursive structure of the transformation matrix of multi-kernel polar codes, like polar codes, permits to drastically reduce the decoding complexity by performing the LLR computation on a kernel base. 
In fact, LLRs can be calculated in the kernel boxes and passed along the Tanner graph of the code to the other kernel boxes from the right to the left, with hard decisions on decoded bits flowing from left to right. 
These hard decisions, representing the estimates of the intermediate bits, are used by kernel boxes to calculate intermediate LLRs. 

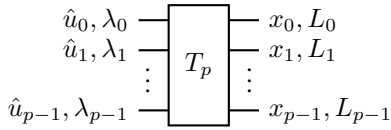
\begin{figure}[htb]
	\begin{center}
		\begin{tikzpicture}[scale=0.2, line width=0.8pt]
		\draw   (0,0) rectangle (4,8) node[midway]{$T_p$};
		\draw[] (-2,7) node[left] {$\hat{u}_0 , \lambda_0$} -- (0,7) ;
		\draw[] (-2,5) node[left] {$\hat{u}_1 , \lambda_1$} -- (0,5) ;
		\draw[] (-0.5,3.5) node[left] {$\vdots$}    (0,3) ;
		\draw[] (-2,1) node[left] {$\hat{u}_{p-1} , \lambda_{p-1}$} -- (0,1) ;
		\draw[] (4,7)                     -- (6,7) node[right] {$x_0 , L_0$} ;
		\draw[] (4,5)                     -- (6,5) node[right] {$x_1 , L_1$} ;
		\draw[] (4,3)                        (4.5,3.5) node[right] {$\vdots$} ;
		\draw[] (4,1)                     -- (6,1) node[right] {$x_{p-1} , L_{p-1}$} ;
		\end{tikzpicture}
	\end{center}
	\caption{A $p \times p$ box corresponding to kernel $T_p$.}
	\label{fig:block-Tl}
\end{figure}

The $p \times p$ box corresponding to a $T_p$ kernel is depicted in Fig.~\ref{fig:block-Tl}, having $u = [u_0,u_1,\ldots,u_{p-1}]$ as input vector and $x = [x_0,x_1,\ldots,x_{p-1}]$ as output vector. 
Hard decisions on the output vector are calculated via multiplication with the kernel matrix, as  $\hat{x} = \hat{u} \cdot T_p$. 
The LLR calculation is more complex; denoting by $L_i$ the input LLRs and by $\lambda_i$ the output LLRs (seen right to left), the SC equation \eqref{SC_eq} can be simplified as
\begin{equation}
\label{f_eq}
\lambda_i = f_i^p( L_0 , L_1 , \dots , L_{p-1} , \hat{u}_0 , \hat{u}_1 , \ldots, \hat{u}_{i-1} ), 
\end{equation}
taking into account only LLRs and bit estimates belonging to the present box. 
The formulation of the LLR update functions $f_i^p$ for specific kernels will be discussed in Sec.~\ref{sec:kernel}.

\section{Kernel Analysis}
\label{sec:kernel}
The polarization phenomenon, originally proved for matrix $T_2$, has been extended to binary matrices in \cite{exp_urbanke} and to arbitrary  finite fields in \cite{MT_qary}, where sufficient and necessary conditions are provided for a square matrix to polarize. 
A polarizing matrix is called \textit{kernel}, and can be used in the construction of polar codes. 
Multi-kernel polar codes are based on the generalized polarization effect obtained mixing kernels of different sizes \cite{MK_pol_proof}. 
Compared to polar codes, our construction permits to exploit the distance properties of the kernels to improve the error correction performance, in particular for small block lengths. 
Here we analyze the structure and the decoding complexity of large kernels, providing recommendations for the design of kernels for multi-kernel polar codes.

%----------------------------------------------
\subsection{Minimum-Distance Spectrum}
%----------------------------------------------

The speed of polarization of a kernel is evaluated via the polarization exponent \cite{non_bin_ker}, calculated through the partial distances of the kernel matrix.  
These partial distances are defined as the minimum weights of a sequence given by the sum of a row and any linear combination of following rows. 
This notion is based on the nature of SC decoding, and is used to drive the kernel design.  
Under SCL decoding, however, polar codes show better performance than predicted by the polarization exponent. 
Moreover, the polarization effect is less important than distance properties for short codes, and kernels should be designed taking this aspect into account.
We conjecture the notion of minimum-distance spectrum \cite{mk_dist_arxiv} to be more effective in these scenarios. 

The \emph{minimum-distance spectrum} $S_{T_p}$ of a kernel $T_p$ is defined as the mapping from dimension $k$, $k=1,2,\ldots,p$, to the largest minimum distance achievable by any $(p,k)$ sub-code of $T_p$, i.e. by any code having $k$ rows of $T_p$ as generator matrix.  
More formally, if $T_p^{(\mathcal{R})}$ is the matrix formed by the rows of $T_p$ indexed by $\mathcal{R} \subset [p]$ and $d(A)$ is the minimum distance of the code generated by the rows of matrix $A$, then the minimum-distance spectrum of $T_p$ is defined as
\begin{equation}
	S_{T_p}(k) = \max_{|\mathcal{R}|=k} d \left( T_P^{(\mathcal{R})} \right) , 
	\label{eq:min-dist-profile}
\end{equation}
for $k=1,\ldots,p$.  
Optimal row set $\mathcal{R}^k_p \subset [p]$ collects the indices of the $k$ rows of $T_p$ forming the generator matrix of the optimal $(p,k,S_{T_p}(k))$ code extracted from the kernel.

The minimum-distance spectrum can be seen as a generalization of partial distances.  
In fact, partial distances are obtained as the distances determined 
by selecting the rows bottom up. 
As a result, row sets are nested, every row set being a subset of all larger ones. 
The minimum-distance spectrum relaxes this constraint, allowing for non-nested information sets, and thus providing a new degree of freedom for the overall code design. 
While partial distances are conceived for characterizing SC decoding, the minimum-distance spectrum seems to be more effective in portraying SCL decoding. 

%----------------------------------------------
\subsection{Kernel Decoding}
%----------------------------------------------

Formulation of decoding equations \eqref{f_eq} for the LLR calculation under SC decoding is of capital importance during kernel design. 
With reference to Fig.~\ref{fig:block-Tl}, the (input) LLRs $L_i$ derive from previous decoding steps or, if the kernel is in the first stage, are the LLRs of bits $x_j$, $L_j = L(x_j)$, calculated from the received symbols; the (output) LLRs $\lambda_i$ represent the LLRs of bits $u_i$, $\lambda_i = L(u_i)$.  They can be expressed using only input LLRs and hard decisions of the previously decoded bits $u_0,\dots,u_{i-1}$, as
\begin{equation}
	\label{eq:gen_lambda}
	\lambda_i = \ln \frac{\sum_{x \in X_0^{(i)}} \exp \left( \sum_{t=0}^{p-1} (1-x_t) L_t \right) }
	                     {\sum_{x \in X_1^{(i)}} \exp \left( \sum_{t=0}^{p-1} (1-x_t) L_t \right) }, 
\end{equation}
$X_a^{(i)} = \{ v \cdot G : \; v = [u_0,\dots,u_{i-1},a,v_{i+1},\dots,v_{p-1}], v_j \in \mathbb{F}_2 \}$, 
$a = 0,1$, which corresponds to the marginalization over the unknown bits $v_{i+1},\ldots,v_{p-1}$ \cite{polar_analysis}.
Since $|X_a^{(i)}|=2^{p-i}$, to compute this expression is in general exponential in the kernel size $p$, even if it can be simplified by human inspection \cite{arb_ker} or trellis-based decoding of block codes \cite{apd}. 
However, this formulation complicates the calculation of input-bit reliability, restricting the design of the code to Monte-Carlo methods.

In practice, analysis of the tree of recurrent relations of the graph inducted by the kernel matrix may permit to discriminate among the different channel observations, rewriting expression \eqref{eq:gen_lambda} using only basic operations of the LLR algebra \cite{LLR_algebra}; 
if bits $x_1,\dots,x_j$ represent a repetition of bit $x_0$ and the corresponding LLRs $L_i$ are based on independent observations, then
\begin{equation}
  L(x_0 | L_0,\dots,L_j) = L_0 + \dots + L_j .
\end{equation}
On the other hand, if $x_0,\dots,x_j$ are independent bits and the LLRs $L_i$ are based on independent observations, then
\begin{equation}
  L(x_0 \oplus \dots \oplus x_j) =  L_0 \boxplus \dots \boxplus L_j 
\end{equation}
where the $\boxplus$ operation is defined as 
\begin{equation*}
  \bplus a_t \triangleq 2 \tanh^{-1} \Bigl( \prod_{t=0}^{j} \tanh \frac{a_t}{2} \Bigr)  \approx \min_{0 \leq t \leq j} (|a_t|) \cdot \prod_{t=0}^{j} \sgn a_t .
\end{equation*}
If it is possible to formulate \eqref{eq:gen_lambda} with an expression involving only $+$ and $\boxplus$ operations of the LLRs $L_0,\dots,L_{p-1}$, we say that the decoding equation is expressed in \emph{reduced form}. 
In such a case, the complexity becomes linear in $p$, and the analysis of the kernel polarization can be simplified as explained in the next sections. 
Reducibility of general decoding equations being an open problem, we conjecture that \eqref{eq:gen_lambda} cannot be expressed in reduced form for all bit positions and all kernels;  
we suggest to approximate irreducible expressions with similar expressions in reduced form. 

%----------------------------------------------
\subsection{Kernel Examples}
\label{sec:kernel-examples}
%----------------------------------------------

Minimum-distance spectrum of the original kernel of size $2$
\begin{equation}
  T_2 = \begin{pmatrix} 1 & 0 \\ 1 & 1 \end{pmatrix} 
  \label{eq:T2}
\end{equation}
is straightforward: for dimension $1$, the second row is used, achieving distance $2$, while for dimension $2$, both rows are used, achieving distance~$1$.  
Thus we have $S_{T_2} = (2,1)$, with optimal row sets $\mathcal{R}_2^1 = \{ 1 \}$ and $\mathcal{R}_2^2 = \{0,1\}$; they are nested since $\mathcal{R}_2^1 \subset \mathcal{R}_2^2$. 
As a proof of concept for the different designs presented in Section~\ref{sec:design}, we introduce kernels 
\begin{equation}
	\label{eq:T3_T5}
  T_3 = \begin{pmatrix} 1 & 1 & 1 \\ 1 & 0 & 1 \\ 0 & 1 & 1 \end{pmatrix} , \quad 
  T_5 = \begin{pmatrix}
              1 & 1 & 1 & 1 & 1 \\ 
              1 & 0 & 0 & 0 & 0 \\
              1 & 0 & 0 & 1 & 0 \\
              1 & 1 & 1 & 0 & 0 \\
              0 & 0 & 1 & 1 & 1
            \end{pmatrix} 
\end{equation}
for the construction of multi-kernel polar codes. 
We calculate their minimum-distance profiles and their decoding equations, showing their flexibility compared to kernels presented in \cite{non_bin_ker}. 

The spectrum of $T_3$ can be calculated as follows.
we select the first row $(1 \; 1 \; 1)$ to maximize the minimum distance, giving minimum distance 3 and $\mathcal{R}_3^1 = \{ 0 \}$; any other row selection would result in a smaller minimum distance, namely 2. 
For a code of dimension $K=2$, the last two rows, $(1 \; 0 \; 1)$ and $(0 \; 1 \; 1)$, are selected, generating a code of minimum distance 2 with $\mathcal{R}_3^2 = \{ 1,2 \}$; any other row selection would result in a smaller minimum distance. 
Finally, the code of dimension $K=3$ requires to select all rows having $\mathcal{R}_3^3 = \{ 0,1,2 \}$, resulting in a code of minimum distance 1. 
The use of not-nested row sets like $\mathcal{R}_3^1 \not\subset \mathcal{R}_3^2$ allows for improved minimum-distance spectrum $S_{T_3}=(3,2,1)$ compared to \cite{non_bin_ker} while keeping the same polarization rate $E=0.42$. 

A similar analysis for $T_5$ leads to the minimum-distance spectrum $S_{T_5}=(5,3,2,1,1)$ with optimal row sets  $\mathcal{R}_5^1 = \{ 0 \}$, $\mathcal{R}_5^2 = \{ 3,4 \}$, $\mathcal{R}_5^3 = \{ 2,3,4 \}$, $\mathcal{R}_5^4 = \{ 1,2,3,4 \}$ and $\mathcal{R}_5^5 = \{ 0,1,2,3,4 \}$. 
On the other side, its polarization rate is $E=0.359$, which is worse than the rate of $0.431$ achieved by optimal kernel in \cite{non_bin_ker}. 
However, the optimal kernel has spectrum $(4,2,2,2,1)$, limiting the achievable minimum distances of the full code to powers of $2$; it is hard to state if proposed $T_5$ spectrum is better, since the full code distance depends on dimension and code length, however it permits a finer quantization in achievable minimum distances. 

Decoding equations in reduced form can be calculated for the presented kernel; the ones for $T_2$ are the well known
\begin{align*}
  f^2_0 : \lambda_0 &=  L_0 \boxplus L_1  , \\
  f^2_1 : \lambda_1 &= (-1)^{\hat{u}_0} \cdot L_0 + L_1  .
\end{align*}
The ones for $T_3$ given in \eqref{eq:T3_T5} are
\begin{align*}
  f^3_0 : \lambda_0 &=  L_0 \boxplus L_1 \boxplus L_2  , \\
  f^3_1 : \lambda_1 &=  (-1)^{\hat{u}_0} \cdot L_0 + L_1 \boxplus L_2  , \\
  f^3_2 : \lambda_2 &=  (-1)^{\hat{u}_0} \cdot L_1 + (-1)^{\hat{u}_0 \oplus \hat{u}_1} \cdot L_2 . 
\end{align*}
Finally, the decoding equations for $T_5$ given in \eqref{eq:T3_T5} are
\begin{align*}
  f^5_0 : \lambda_0 &= L_1 \boxplus L_2 \boxplus L_4  , \\[1ex]
  f^5_1 : \lambda_1 &= (-1)^{\hat{u}_0} \cdot ( L_0 \boxplus L_3 \boxplus ( L_2 + ( L_1 \boxplus L_4 ) ) )  , \\[1ex]
  f^5_2 : \lambda_2 &= (-1)^{\hat{u}_1 } \cdot ( L_0 \boxplus L_1 )  + ( L_3 \boxplus L_4 )  , \\[1ex]
f^5_3 : \lambda_3 &= (-1)^{\hat{u}_0 \oplus \hat{u}_1 \oplus \hat{u}_2} \cdot L_0 + (-1)^{\hat{u}_0} \cdot L_1  + ( L_2 \boxplus ( L_3 + L_4) )  , \\[1ex]
f^5_4 : \lambda_4 &= (-1)^{\hat{u}_0 \oplus \hat{u}_3} \cdot L_2 + (-1)^{\hat{u}_0 \oplus \hat{u}_2} \cdot L_3   + (-1)^{\hat{u}_0} \cdot L_4 .
\end{align*}
All expressions are optimal apart from $f_2^5$. 
Conjecturing the original expression to be irreducible, we approximate it with a reducible one. 
This partially effects the decoding performance, however it allows for analytical determination of reliabilities to be used for code design as shown in Section~\ref{sec:reliability}.

\section{Code Design}
\label{sec:design}
Multi-kernel polar codes introduce new options in the code design, which for the original polar codes is limited to the selection of the information set according to reliabilites. 
In this section, we describe three design principles for multi-kernel polar codes, called the \emph{reliability} design, the \emph{distance} design and the \emph{hybrid} design, theoretically motivating them and providing for each one a practical design algorithm. 

%--------------------------------------------------------
\subsection{Reliability Design}
\label{sec:reliability}
%--------------------------------------------------------

The reliability design is based on the polarization phenomenon and aims at minimizing the probability of error under SC decoding. 
This design is conceived for long codes, where channel polarization is strong enough to discriminate the channels properly. 
After a brief review of the concept, we will show how to determine these reliabilities, concluding the section with a discussion on the optimal kernels order.

We upper-bound the error probability under SC decoding of an $(N,K)$ multi-kernel polar code with information set $\mathcal{I}$ by
\begin{equation}
	P_e^\mathsf{SC} \le \sum_{i \in \Iset} P_e(u_i) ,
	\label{eq:SC-word-error-upper-bound}
\end{equation}
where $P_e(u_i) = P( \hat{u}_i \ne u_i | \hat{u}_j = u_j, j < i )$ denotes the probability of making a wrong decision for bit $\hat{u}_i$ assuming that all previously decoded bits are correct. 
The reliability design aims to minimize this upper bound so that the information set $\Iset^\mathsf{R}$ is chosen to contain the $K$ most reliable positions.  
This is equivalent to finding the $K$ positions minimizing the maximum error probability within these positions; $\Iset^\mathsf{R}$ can hence be found as the solution of the optimization problem 
\begin{align}
\label{eq:inf-set-RBD-prob}
\min \quad & \max_{i \in \Iset} P_e(u_i) \\
\text{s.t.} \quad & \Iset \subset [N] , |\Iset| = K. \notag 
\end{align}

The simplest way to calculate the reliabilities is to use Monte-Carlo simulation, e.g. to run a genie-aided SC decoder to estimate the error rate of each input bit.  
In more detail, the all-zero codeword is transmitted over a channel with a target design SNR, and is decoded with a modified SC decoder that counts bit errors based on hard decisions of the LLRs but feeds back the correct decisions. 
As this method requires a large number of simulations to get stable results, we suggest to compute the approximated reliabilities of the input bits by density evolution under Gaussian approximation (DE/GA) \cite{DE_mori} instead. 
If we suppose LLR distributions to be Gaussian, their variance is twice their mean value, i.e., $L_i \sim \mathcal{N}(m_i, 2 m_i)$, permitting to follow their evolution by tracking their mean. 

For DE/GA, the means are passed in the Tanner graph from the right to the left, similarly to LLRs. 
Looking at the Tanner graph block depicted in Fig.~\ref{fig:block-Tl}, we denote the mean of $\lambda_i$ by $\mu_i$, and the mean of $L_j$ by $m_j$.  
For a BI-AWGN transmission system, the initial channel LLRs are distributed as $l_i \sim \mathcal{N} \left( \frac{2}{\sigma^2},\frac{4}{\sigma^2} \right) $, the initial mean value being $m_i = \frac{2}{\sigma^2}$ \cite{LDPC_analysis}.
Under the Gaussian assumption, the error probability $P_e(u_i)$ is in direct correspondence with the LLR mean value $\mu_i$ as $P_e(u_i) = Q(\sqrt{\mu_i/2})$, where $Q(.)$ denotes the tail probability of the standard Gaussian distribution; hence the SC error probability can be lower bounded by  
\begin{equation}
	P_e^\mathsf{SC} \ge \max_{i \in \Iset} Q(\sqrt{\mu_i/2})  .
	\label{eq:SC-word-error-lower-bound-Q}
\end{equation}
To reduce the complexity of \eqref{eq:inf-set-RBD-prob}, error probabilities may be replaced LLR mean values; $\Iset^\mathsf{R}$ can then be determined solving 
\begin{align}
\label{eq:inf-set-RBD-mean}
\max \quad & \min_{i \in \Iset} \mu_i \\
\text{s.t.} \quad & \Iset \subset [N] , |\Iset| = K, \notag 
\end{align}
where $\mu_i$ denotes the mean value of the LLR of $u_i$.

If kernel decoding equations \eqref{f_eq} are expressed in reduced form, then the equations tracking the LLR means can be written directly: if LLRs $L_0,\dots,L_{j-1}$ are independent, then
\begin{align}
  \mu(L_0 + \dots + L_{j-1}) &=  m_0 + \dots + m_{j-1}  \\
  \mu(L_0 \boxplus \dots \boxplus L_{j-1}) &=  \varphi_{j} ( m_0, \dots, m_{j-1} ),
\end{align} 
where 
\begin{equation}
\varphi_{j}( m_0, \dots, m_{j-1} ) = \phi^{-1} \left( 1 - \prod_{t=0}^{j-1} (1-\phi(m_t)) \right),
\end{equation}
\begin{equation}
\phi(m) = 1 - \frac{1}{\sqrt{4 \pi m}} \int_{-\infty}^{+\infty} \tanh \frac{u}{2} e^{-\frac{(u-m)^2}{4m}} du .
\end{equation}
We recall that functions $\phi$ and $\phi^{-1}$ can be approximated as 
\begin{align}
\phi(m) \approx \left\{
\begin{array}{l l}   
 e^{a m^2 - b m} & \text{if } 0 \leq m < c   \\
 e^{-\alpha m^{\gamma} + \beta} & \text{if } m \geq c 
\end{array} 
\right.
\label{equ:psi} \\
\phi^{-1}(m) \approx \left\{
\begin{array}{l l}   
 \frac{b - \sqrt{b^2+4a\ln{m}}}{2a} & \text{if } 0 \leq m < c   \\
 \left( \frac{\beta - \ln{m}}{\alpha} \right)^{\frac{1}{\gamma}} & \text{if } m \geq c 
\end{array} 
\right.
\label{equ:psi_inv}
\end{align}
Parameters $\alpha = 0.4527$, $\beta = 0.0218$, $\gamma = 0.86$, $a = 0.0564$, $b = 0.48560$, $c = 0.867861$ are acquired by curve-fitting \cite{psi_approx}. 

The decoding equations for the kernels presented in Section \ref{sec:model} lead to the following evolution of mean values. 
\begin{description}
\item[DE/GA for kernel $T_2$:]
  \begin{align*}
    \mu_0 &= \varphi_2( m_0,m_1 ) \\
    \mu_1 &= m_0 + m_1 
  \end{align*}  
\item[DE/GA for kernel $T_3$:]
  \begin{align*}
    \mu_0 &= \varphi_3( m_0,m_1,m_2 ) \\
    \mu_1 &= m_0 + \varphi_2( m_1,m_2 ) \\
    \mu_2 &= m_1 + m_2 
  \end{align*}
\item[DE/GA for kernel $T_5$:]
  \begin{align*}
    \mu_0 &= \varphi_3(m_1,m_2,m_4) \\
    \mu_1 &= \varphi_3(m_0,m_3,(m_2 + \varphi_2(m_1,m_4))) \\
    \mu_2 &= \varphi_2(m_0,m_1) + \varphi_2(m_4,m_4) \\
    \mu_3 &= m_0 + m_1 + \varphi_2(m_2,m_3+m_4) \\
    \mu_4 &= m_2 + m_3 + m_4 
  \end{align*}
\end{description}

The last point to be addressed is the selection of the order of kernels.  
The kernel order has two aspects.  
First, transformation matrices obtained by permuting the same kernels lead to equivalent codes by conveniently selecting the two information sets due to the permutation property of the Kronecker product.  
However, it is hard to predict the impact of the kernel order on the polarization of the input bits due to the non-linearity of function $\phi$.  
Therefore codes of same length and dimension but different kernel orders may be different if reliability design is performed; as a result, for specific code dimensions, one kernel order may be preferable to others. 
We propose to perform an exhaustive search among all possible kernel orders to find the best one.  
For a code of dimension $K$, the metric used for the order selection is the sum of the reliabilities of the best $K$ bits, and the kernel order giving the largest sum is retained.

%--------------------------------------------------------
\subsection{Distance Design}
\label{sec:distance}
%--------------------------------------------------------

In this section, we describe a design for multi-kernel polar codes maximizing the minimum distance of the resulting code \cite{mk_dist_arxiv}. 
This design is envisaged for short codes, where the polarization effect is not strong enough to prevail over the minimum distance properties. 
For an $(N,K)$ multi-kernel polar code, the probability of error under maximum-likelihood (ML) decoding for the AWGN channel is bounded below as
\begin{equation}
	P_e^\mathsf{ML} \ge Q(\sqrt{d \mu / 2}) ,
	\label{eq:ML-word-error-lower-bound}
\end{equation}
where $d$ denotes the minimum distance of the code and $\mu = 2/\sigma^2$ the mean of the channel LLR.  
If the information set $\Iset^\mathsf{D}$ is selected to maximize $d \mu$, then \eqref{eq:ML-word-error-lower-bound} is minimized; since $\mu$ is fixed, this corresponds to solving the optimization problem 
\begin{align}
\label{eq:inf-set-DBD}
\max \quad &  d_N(\Iset) \\
\text{s.t.} \quad & \Iset \subset [N] , |\Iset| = K, \notag 
\end{align}
where $d_N(\Iset)$ denotes the minimum distance of the code defined by the rows of $T_N$ indexed by~$\Iset$.
We solve this problem in two steps: first, we calculate the optimal minimum distance through the minimum distance spectrum of $T_N$, then we find the information set achieving that distance. 
\begin{algorithm}[tbh]
	\caption{Information set for minimum distance} 
	\label{alg:min-dist-design}
	\begin{algorithmic}[1]
		\State $\text{Initialize the sets } \mathcal{I} = \emptyset \text{ and } \mathcal{R}_p^0  = \emptyset$
		\State $\text{Load vector } s_N = (2,1)^{\otimes n} \otimes S_{T_p}$
		\State $\text{Load optimal row sets } \mathcal{R}_p^1,\dots,\mathcal{R}_p^p$
		\For{$k = 1 \dots K$}
   			\State $l = \text{argmax} (s_N)$
   			\State $c = (l \: \text{mod} \: p)$
   			\State $q = \lfloor \frac{N-l-1}{p} \rfloor$ \label{alg:q}
   			\State $\mathcal{I} = \left( \mathcal{I} \setminus (\mathcal{R}_p^{c} + qp) \right) \cup (\mathcal{R}_p^{c+1} + qp)$ \label{alg:I}
   			\State $s_N(l) = 0$
		\EndFor
		\State\Return $\mathcal{I}$ 
	\end{algorithmic}
\end{algorithm}

Though finding the minimum-distance spectrum of a code is in general a complex task, for polar codes it can be easily computed as $S_{T_2^{\otimes n}} = \sort([2 \quad 1]^{\otimes n})$, where the vector is sorted in descending order \cite{pol_dist}. 
This property can be generalized to multi-kernel polar codes as follows. 
\begin{prop}
\label{prop:spectrum} 
If $T_N = T_2^{\otimes n} \otimes T_p$, then 
\begin{equation}
	S_{T_N} = \sort(S_{T_2^{\otimes n}} \otimes S_{T_p}) = \sort( [2 \quad 1]^{\otimes n} \otimes S_{T_p} ). 
	\label{eq:min-dist-profile-mk-codes}
\end{equation}
\begin{proof}
The property obviously holds for $n=0$. 
By inductive hypothesis we now suppose that it holds for $n-1$, i.e., that $S_{T_{N/2}} = \sort(S_{T_2^{\otimes n-1}} \otimes S_{T_p})$. 
Defining $a^U = S_{T_{N/2}}$, $a^L = 2 S_{T_{N/2}}$, such that $a = \sort([a^U,a^L])$, the proposition is proved if $a = S_{T_N}$ since
\begin{align*}
	\sort( S_{T_{N/2}} , 2 S_{T_{N/2}} )
	&= \sort( [2 \quad 1] \otimes  S_{T_{N/2}} ]) =  \\
	&= \sort( [2 \quad 1] \otimes  \sort( [1 , 2] \otimes  S_{T_{N/4}} ) ) =  \\
	&= \sort( [2 \quad 1] \otimes  [1 , 2] \otimes  S_{T_{N/4}} ) = \ldots =  \\
	&= \sort( [2 \quad 1]^{\otimes n} \otimes S_{T_p} ).
\end{align*}  
Consider a subcode of $T_N = \tiny\begin{pmatrix} T_{N/2} & 0 \\ T_{N/2} & T_{N/2} \end{pmatrix}$ defined by $K$ rows, with $K^U$ rows from $T^U = \begin{bmatrix} T_{N/2} & 0 \end{bmatrix}$, $K^L$ rows from $T^L = \begin{bmatrix} T_{N/2} & T_{N/2} \end{bmatrix}$, and $K^U + K^L = K$; denote the corresponding submatrices of $T^U$ and $T^L$ by $T^U_A$ and $T^L_B$, respectively. 
If row indices are selected such that $d(T^U_A) = a^U(K^U)$ and $d(T^L_B) = a^L(K^L)$, which is possible by the induction hypotheses, then the minimum distance of this subcode is
\begin{align*}
	d \Bigl( \begin{bmatrix} T^U_A \\ T^L_B \end{bmatrix} \Bigr)
	&\overset{(a)}{=}   \min \bigl( d(T^U_A) , d(T^L_B) \bigr) = \\
	&= \min \bigl( a^U(K^U) , a^L(K^L) \bigr)
	\overset{(b)}{=} a(K) 
\end{align*}
where (a) follows from the distance property of the $(u | u+v)$ construction \cite{McW-S} and (b) from the sorting of two sorted lists.
\end{proof} 
\end{prop}
Proposition~\ref{prop:spectrum} requires the transformation matrix of the code to be in the form $T_N = T_2^{\otimes n} \otimes T_p$. 
Note that this is not a limiting construction for the minimum-distance construction, since the minimum-distance spectrum does not depend on the order of the kernels in the Kronecker product. 
However, this structure permits to divide $T_N$ into $2^n$ sub-matrices of $p$ rows, termed as \textit{sectors} in the following, each one consisting of a vector of $T_p$ kernels and all-zero matrices.
Analogously, information sets $\Iset$ of size $K$ can be split into $2^n$ smaller information sets $\Iset_q \subseteq [p]$, $|\Iset_q|=K_q\leq p$ and $\sum_q K_q = K$, where each $\Iset_q$ collects the rows of the $q$-th sector included in $\Iset$. 
Since the minimum distance spectrum of $q$-th sector is given by $S_q = 2^{wt(q)} \cdot S_{T_p}$, where $wt(q)$ is the number of ones of $q$-th row of $T_2^{\otimes n}$, this division permits to identify the contribution of each sector to the minimum distance of the code. 
Due to the distance property of the $(u | u+v)$ construction, $d_N(\Iset) = \min \left( d_p(\Iset_1),\dots,d_p(\Iset_{2^n}) \right)$; if $\Iset_q$ is formed by the optimal row set $\mathcal{R}_p^{K_q}$, then $d_p(\Iset_q) = S_q(K_q)$. 

This concept is exploited in greedy Algorithm~\ref{alg:min-dist-design} to design multi-kernel polar codes with optimal minimum distance. 
The algorithm adds sequentially row indices to the information set $\mathcal{I}$, modifying the information set of dimension $K-1$ to obtain the one for dimension $K$. 
Vector $s_N = [S_1|\dots|S_{2^n}]$ formed by the minimum-distance spectra of the individual sectors is initially calculated as $s_N = (2,1)^{\otimes n} \otimes S_{T_p}$; note that the vector is not sorted.  
At step $k$, the position $l$ of the largest entry in $s_N$ is extracted as $b=s_N(l)$, and $s_N(l)$ is set to zero. 
Then $b$ represents the best minimum distance achievable by the code for dimension $K$; $q$ defined in line~\ref{alg:q} identifies the sector to be updated to reach that distance; index $c = l \mod p$ represents the number of rows of sector $q$ already included in $\Iset$. 
To increase the value of $c$ by one, the algorithm substitutes the previous optimal row set $\mathcal{R}^{c}_p$ with the following one, i.e. $\mathcal{R}_p^{c+1}$, in line~\ref{alg:I}. 
The optimal row sets of the individual sectors need to be shifted by $qp$ to be properly included in $\Iset$. 
The algorithm stops when $\Iset$ comprises $K$ elements. 

\begin{algorithm}[tbh]
\caption{Spectrum of Kronecker product of kernels} 
\label{algo_2}
\label{alg:spectrum-kronecker-product}
\begin{algorithmic}[1]
\State $\text{Load optimal row sets } \mathcal{R}_{p_1}^1,\dots,\mathcal{R}_{p_1}^{p_1}, \mathcal{R}_{p_2}^1,\dots,\mathcal{R}_{p_2}^{p_2}$
\For{$k = 1 \dots p$}
   \State $S_{T_p}(k) = 0$
   \State $\kappa = \text{ListPartition}(k,p_1,p_2)$
   \For{$\ell = 1 \dots \text{length}(\kappa)$}
      \State $\langle k_1,\dots,k_t \rangle = \kappa(\ell)$
      \State $\mathcal{R} = \bigcup_{j=1}^{t} \left(\mathcal{R}_{p_2}^{k_j} + \mathcal{R}_{p_1}^t(j) \cdot p_2 \right)$
      \State $m = \text{MinDist}(T_p(\mathcal{R},:))$
      \If{$m > S_{T_p}(k)$}
         \State $S_{T_p}(k) = m$
         \State $\mathcal{R}_p^k = \mathcal{R}$
      \EndIf
   \EndFor
\EndFor
\State\Return $\mathcal{R}_p^1,\dots,\mathcal{R}_p^p,S_{T_p}$ 
\end{algorithmic}
\end{algorithm}

Algorithm~\ref{alg:min-dist-design} requires the minimum-distance spectrum $S_{T_p}$ and the optimal row sets of kernel $T_p$. 
A brute force calculation may be prohibitive for large kernels, since it is required to check the distances generated by all the $\binom{p}{k}$ possible $k$-rows sub-matrices of $T_p$.  
However, if $T_p = T_{p_1} \otimes T_{p_2}$, the kernel $T_p$ can be divided into $p_1$ sectors of $p_2$ rows, each one formed by the juxtaposition of kernels $T_{p_2}$. 
Then the set of $k$ row indices, given by $\mathcal{R}^k_p$, is partitioned according to the sectors, indexed by the set $\{i_1,...,i_t\}$, and within each sector there are $k_j$ rows, $j=1,\ldots,t$, where $k = \sum_{j=1}^t k_j$. 
Given this structure, we propose to limit the search space for the sector indices and for the index sets within the sectors to optimal row sets of the component kernels.

In more detail, an optimal row set $\mathcal{R}_{p_1}^{t} = \{i_1,\dots,i_t\}$ identifies the indices of the sectors that will contribute to $\mathcal{R}^k_p$; for every retained sector $i_j$, an optimal row set $\mathcal{R}_{p_2}^{k_j}$ is included in $\mathcal{R}^k_p$. 
For each $k$, all the possible combinations of $t$ and $k_j$ have to be checked. 
Algorithm~\ref{alg:spectrum-kronecker-product} performs this task, comparing the minimum distances of the row sets generated by all integer partitions of $k$ of maximum length $p_1$, i.e. the set of all possible ways of writing $k$ as a sum of up to $p_1$ positive integers not larger than $p_2$.
The integer partition $\langle k_1,\dots,k_t \rangle$, where $k = \sum_{i=1}^t k_i$, unambiguously identifies row set
\begin{equation}
\label{eq:p1p2}
\mathcal{R}_{\langle k_1,\dots,k_t \rangle} = \bigcup_{j=1}^{t} \left( \mathcal{R}_{p_2}^{k_j} + i_j \cdot p_2 \right),
\end{equation}
where $\mathcal{R}_{p_1}^{t} = \{i_1,\dots,i_t\}$ with $i_1 < \dots < i_t$. 
In practice, $t$ sectors are included in the row set, whose indices are listed in $\mathcal{R}_{p_1}^{t}$. 
The $j$-th sector contributes with $k_j$ rows, that are chosen according to the optimal row set of kernel $T_{p_2}$.  

As an example, we show the steps performed by Algorithm~\ref{alg:spectrum-kronecker-product} to compute the optimal row set $\mathcal{R}^4_9$ for kernel
\begin{equation*}
\label{eq:G_9_TM}
T_{9} = T_3 \otimes T_3 = \left( 
\begin{array}{ c c c|c c c|c c c}
 1 & 1 & 1 & 1 & 1 & 1 & 1 & 1 & 1 \\
 \hdashline
 1 & 0 & 1 & 1 & 0 & 1 & 1 & 0 & 1 \\
 \hdashline
 0 & 1 & 1 & 0 & 1 & 1 & 0 & 1 & 1 \\
 \hline
 1 & 1 & 1 & 0 & 0 & 0 & 1 & 1 & 1 \\
 \hdashline
 1 & 0 & 1 & 0 & 0 & 0 & 1 & 0 & 1 \\
 \hdashline
 0 & 1 & 1 & 0 & 0 & 0 & 0 & 1 & 1 \\
 \hline
 0 & 0 & 0 & 1 & 1 & 1 & 1 & 1 & 1 \\
 \hdashline
 0 & 0 & 0 & 1 & 0 & 1 & 1 & 0 & 1 \\
 \hdashline
 0 & 0 & 0 & 0 & 1 & 1 & 0 & 1 & 1 \\
\end{array}
\right) .
\end{equation*}
In this case, only $3$ integer partitions of $k = 4$ respect the required properties, namely $\langle 1,3 \rangle$, $\langle 2,2 \rangle$ and $\langle 1,1,2 \rangle$. 
The optimal row sets associated to $\langle 1,3 \rangle$ is given by \eqref{eq:p1p2} as
\begin{equation}
\mathcal{R}_{\langle 1,3 \rangle} = \left( \mathcal{R}_{p_2}^{1} + 1 \cdot 3 \right) \cup \left( \mathcal{R}_{p_2}^{3} + 2 \cdot 3 \right) = \{ 3,6,7,8 \}\end{equation}
with minimum distance $\text{MinDist}(T_9(\mathcal{R}_{\langle 1,3 \rangle},:)) = 2$; similarly, Algorithm~\ref{alg:spectrum-kronecker-product} calculates $\mathcal{R}_{\langle 2,2 \rangle} = \{ 4,5,7,8 \} $ with minimum distance $4$ and $\mathcal{R}_{\langle 1,1,2 \rangle} = \{ 0,3,7,8 \} $ with minimum distance $3$. 
Algorithm~\ref{alg:spectrum-kronecker-product} selects $\mathcal{R}_9^4 = \mathcal{R}_{\langle 2,2 \rangle} = \{ 4,5,7,8 \}$, with $S_{T_9}(4) = 4$. 
The complete minimum-distance spectrum of $T_9$ obtained by running Algorithm~\ref{alg:spectrum-kronecker-product} for $k=1,\dots,9$ is $S_{T_9} = (9,6,4,4,3,2,2,2,1)$. 
Note that this spectrum is optimal, as we verified by an exhaustive search. 
In general the spectrum achieved by Algorithm~\ref{alg:spectrum-kronecker-product} may be suboptimal.

%--------------------------------------------------------
\subsection{Hybrid Design}
\label{sec:hybrid}
%--------------------------------------------------------

The reliability design is conceived for SC decoding, and therefore suited for very long codes, where SC decoding becomes asymptotically optimal.  
The distance design, on the other hand, is assuming ML decoding, and thus is suited for short codes under SCL decoding, where moderate list lengths approximate ML decoding very well.  
The hybrid design combines reliability and distance as design criteria, and it is particularly effective to construct multi-kernel polar codes for medium code lengths under SCL decoding.  

\begin{figure}[tb]
    \center
	\includegraphics[width=0.40\textwidth]{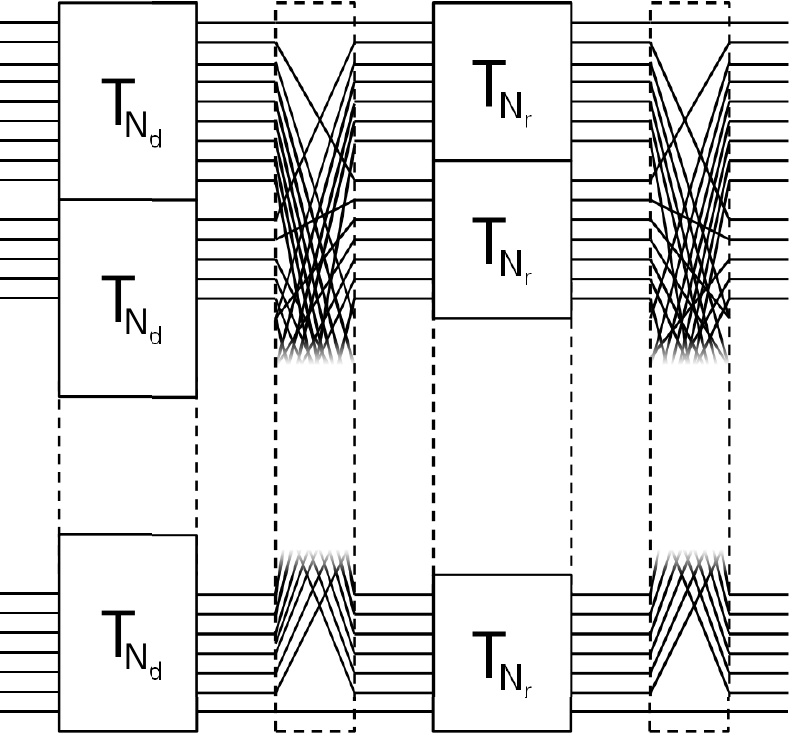}
	\caption{Tanner graph of $T_N = T_{N_r} \otimes T_{N_d}$ for the hybrid design.}
	\label{fig:tanner-graph-hybrid}
\end{figure}

To introduce the hybrid design principle, we partition the transformation matrix \eqref{equ:GN-MK} of a multi-kernel polar code as 
\begin{equation}
	T_N = T_{N_r} \otimes T_{N_d} ,
	\label{equ:transform-hybrid-design}
\end{equation}
with $T_{N_r} = T_{p_1} \otimes \dots \otimes T_{p_\psi}$ and $T_{N_d} = T_{p_{\psi+1}} \otimes \dots \otimes T_{p_s}$.  
The two matrices $T_{N_r}$ and $T_{N_d}$ can be treated as transformation matrices of smaller multi-kernel polar codes, of length $N_r = p_1 \cdot \dots \cdot p_\psi$ and $N_d = p_{\psi+1} \cdot \dots \cdot p_s$ respectively, with $N = N_r \cdot N_d$; corresponding Tanner graph is depicted in Fig.~\ref{fig:tanner-graph-hybrid}.
The idea is to apply the distance design to the left part of the graph, consisting of $T_{N_d}$ blocks, and the reliability design to the right part of the graph, consisting of $T_{N_r}$ blocks; indices `$d$' and `$r$' stand for distance and reliability, respectively. 
This hybrid design comprises a parameter, $\psi$, that allows to trade distance vs.\ reliability. 
Reliability and distance designs can be seen as extreme cases with $\psi = s$ and $\psi = 0$ respectively.   

Consider now the following decoding principle.  
The normal SC decoder proceeds until all right-messages are available at the input to the first $T_{N_d}$ block.  
The block makes a local ML decision, i.e., it decides for the most likely codeword, and this decision is fed back into the SC decoding process.  
SC decoding proceeds until all right-messages are available at the input to the second $T_{N_d}$ block, which makes a local ML decision and feeds the result back into the decoding process.  
This continues until the last $T_{N_d}$ block has made its decision.  
Note that this decoding principle can be approximated by a plain SCL decoder, where larger values of $\psi$ may require larger list sizes to reach ML decoding of left blocks.

Consider now the probability of error of the local ML decision at the $i$-th $T_{N_d}$ block, denoted by $P_e(\boldsymbol{u}_i)$ for any $i=0,1,\ldots,N_r-1$, assuming all previous decisions being error-free.  
By SC decoding, all incoming messages have the same reliability; imposing Gaussian approximation on the message densities, we denote the mean of the incoming message density by $\mu_i$ and the minimum distance of the local code, as imposed by the information set on the $T_{N_d}$ block, by $d_i$.  
The probability of error can thus be lower-bounded by
\begin{equation}
	P_e(\boldsymbol{u}_i) \ge Q(\sqrt{d_i \mu_i / 2}) ,
	\label{equ:local-ML-word-error-prob}
\end{equation}
while for the overall decoder is lower-bounded as
\begin{equation}
	P_e^\mathsf{MLSC} \ge \max_{i \in [N_r]} P_e(\boldsymbol{u}_i) \ge \max_{i \in [N_r]} Q(\sqrt{d_i \mu_i / 2}) .
	\label{eq:MLSC-word-error-lower-bound}
\end{equation}
The information set $\Iset^\mathsf{H}$ of the hybrid design is selected to minimize this lower-bound, namely solving the optimization problem maximizing the minimum of the terms $d_i \mu_i$ as
\begin{align}
\label{eq:inf-set-DRBD}
\max \quad & \min_{i \in [N_r]} d_i(\Iset) \mu_i \\
\text{s.t.} \quad & \Iset \subset [N] , |\Iset| = K, \notag 
\end{align}
where $d_i(\Iset)$ denotes the minimum distance of the code induced by $\Iset$ over the $i$-th $T_{N_d}$ block. 
This hybrid design minimizes the error rate for the mixed ML-SC decoder as described above. 
Note that both the bound \eqref{eq:MLSC-word-error-lower-bound} and the information set \eqref{eq:inf-set-DRBD} are for a fixed value of $\psi \in [0,s]$, which may be adapted to the available list length of the SCL decoder. 

The optimisation for the information set, as given in \eqref{eq:inf-set-DRBD}, can be solved slightly modifying Algorithm~\ref{alg:min-dist-design}, introduced for the distance design in the previous section.   
Initially, the reliabilities of the $N_r$ input bits of the partial transformation matrix $T_{N_r}$ are determined using DE/GA, as described in Section~\ref{sec:reliability}.  These reliabilities are stored in an intermediate vector $\mu = (\mu_{N_r-1}, \ldots, \mu_0)$, 
where $\mu_i$ represents the reliability of the $i$-th input bit of the code generated by $T_{N_r}$.
At the same time, the minimum distance spectrum of the partial transformation matrix $T_{N_d}$ is computed, along with the optimal rows sets, using the methods for the distance design, as described in Section~\ref{sec:distance}.
Finally, vector $d_N = \mu \otimes S_{T_{N_d}}$, representing the "hybrid" spectrum of $T_N$, is calculated and given to Algorithm~\ref{alg:min-dist-design}, along with the optimal rows sets calculated previously, to design the information set of the code.  

%--------------------------------------------------------
\subsection{Construction Example}
\label{subsec:example}
%--------------------------------------------------------

We illustrate the proposed designs through a multi-kernel polar code of length $N = 12$ and dimension $K = 4$ with transformation matrix $T_{12} = T_2^{\otimes 2} \otimes T_3$ depicted in Fig.~\ref{fig:G_12}, i.e. 
\begin{equation*}
\label{eq:G_12_TM}
T_{12} = \left( 
\begin{array}{ c c c:c c c|c c c:c c c}
 1 & 1 & 1 & 0 & 0 & 0 & 0 & 0 & 0 & 0 & 0 & 0 \\
 1 & 0 & 1 & 0 & 0 & 0 & 0 & 0 & 0 & 0 & 0 & 0 \\
 0 & 1 & 1 & 0 & 0 & 0 & 0 & 0 & 0 & 0 & 0 & 0 \\
 \hdashline
 1 & 1 & 1 & 1 & 1 & 1 & 0 & 0 & 0 & 0 & 0 & 0 \\
 1 & 0 & 1 & 1 & 0 & 1 & 0 & 0 & 0 & 0 & 0 & 0 \\
 0 & 1 & 1 & 0 & 1 & 1 & 0 & 0 & 0 & 0 & 0 & 0 \\
 \hline
 1 & 1 & 1 & 0 & 0 & 0 & 1 & 1 & 1 & 0 & 0 & 0 \\
 1 & 0 & 1 & 0 & 0 & 0 & 1 & 0 & 1 & 0 & 0 & 0 \\
 0 & 1 & 1 & 0 & 0 & 0 & 0 & 1 & 1 & 0 & 0 & 0 \\
 \hdashline
 1 & 1 & 1 & 1 & 1 & 1 & 1 & 1 & 1 & 1 & 1 & 1 \\
 1 & 0 & 1 & 1 & 0 & 1 & 1 & 0 & 1 & 1 & 0 & 1 \\
 0 & 1 & 1 & 0 & 1 & 1 & 0 & 1 & 1 & 0 & 1 & 1 
\end{array}
\right) ,
\end{equation*}
for a BI-AWGN channel with inputs $\pm 1$ and $\sigma^2 = 0.5$.
\subsubsection{Reliability design}
Using DE/GA, the reliabilities are calculated as $(0.09,$ $1.28,$ $2,$ $1.85,$ $7.3,$ $9.12,$ $2.75,$ $9.57,$ $11.56,$ $11.94,$ $29.42,$ $32)$.
The $K = 4$ most reliable positions form the information set: $\Iset^\mathsf{R} = \{ 8,9,10,11 \}$. 
\subsubsection{Distance design}
The minimum-distance spectrum and the optimal row sets of the kernel $T_3$ depicted in \eqref{eq:T3_T5} are $S_{T_3} = (3,2,1)$ with $\mathcal{R}_3^1 = \{0\}$, $\mathcal{R}_3^2 = \{1,2\}$ and $\mathcal{R}_3^3 = \{0,1,2\}$; the auxiliary vector is calculated as $d_{12} = (2,1)^{\otimes 2} \otimes (3,2,1) = ( 12,$ $8,$ $4,$ $6,$ $4,$ $2,$ $6,$ $4,$ $2,$ $3,$ $2,$ $1)$. 
Algorithm~\ref{alg:min-dist-design} for $K = 4$ gives then $\Iset^\mathsf{D} = \{ 3 , 6 , 10 , 11 \}$. 
\subsubsection{Hybrid design}
Transformation matrix $T_{12}$ includes $s=3$ kernels, thus $\psi \in \{0,1,2,3\}$.  
The hybrid design results in the reliability design for $\psi = 3$ and in the distance design for $\psi = 0$. 
For $\psi = 2$, we have $T_{N_r} = T_2^{\otimes 2}$ and $T_{N_d} = T_3$. 
The reliabilities of $T_{N_r}$ are determined by DE/GA, resulting in $\mu = (16,$ $5.78,$ $4.56,$ $1)$, while $S_{T_3}$ has been determined above.  
Thus we obtain the mixed spectrum $s_{12} = \mu \otimes S_{T_3} = (48,$ $32,$ $16,$ $17.34,$ $11.56,$ $5.78,$ $13.68,$ $9.12,$ $4.56,$ $3,$ $2,$ $1 )$
for the transformation matrix $T_{12}$. 
This vector along with the optimal row sets for $T_3$ is used as input to Algorithm~\ref{alg:min-dist-design}, giving the information set $\mathcal{I}^\mathsf{H} = \{ 6,9,10,11 \}$ for $K = 4$. 
For $\psi = 1$ instead, we have $T_{N_r} = T_2$ and $T_{N_d} = T_2 \otimes T_3$, with vector $\mu = (8,2.28)$. 
The minimum-distance spectrum of $T_{N_d}$ can be calculated by Algorithm~\ref{alg:spectrum-kronecker-product} as $S_{T_2 \otimes T_3} = (6,4,3,2,2,1)$, with $\mathcal{R}_6^1 = \{3\}$, $\mathcal{R}_6^2 = \{4,5\}$, $\mathcal{R}_6^3 = \{0,4,5\}$, $\mathcal{R}_6^4 = \{0,3,4,5\}$, $\mathcal{R}_6^5 = \{1,2,3,4,5\}$ and $\mathcal{R}_6^6 = \{0,1,2,3,4,5\}$.
Algorithm~\ref{alg:min-dist-design} takes then as inputs $d_{12} = \mu \otimes S_{G_d} = ( 48 ,$ $32 ,$ $16 ,$ $24 ,$ $ 16 ,$ $8 ,$ $13.68 ,$ $9.12 ,$ $4.56 ,$ $6.84 ,$ $4.56 ,$ $2.28 )$ 
and provides the information set $\mathcal{I}^\mathsf{H} = \{ 6 , 9 , 10 , 11 \}$. 

The four designs are summarized in Table~\ref{table:example-design-comparison}.  
As expected, distance design leads to the best minimum distance of~6, while the other designs yield a minimum distance of 4.  
In this case, $\psi = 1,2$ result in the same information set, which is however different from the reliability design.  

\begin{table}[h]
	\begin{center}
		\def\arraystretch{1.5}
			\begin{tabular}{| c | c | c |}
				\hline
				$\psi$    & $\mathcal{I}$           & min. dist. \\ \hline
				3 (rel.)  & $\{ 8 , 9 , 10 , 11 \}$ & 4          \\ \hline
				2         & $\{ 6 , 9 , 10 , 11 \}$ & 4          \\ \hline
				1         & $\{ 6 , 9 , 10 , 11 \}$ & 4          \\ \hline
				0 (dist.) & $\{ 3 , 6 , 10 , 11 \}$ & 6          \\ \hline
			\end{tabular}
	\end{center}
	\caption[]{$\mathcal{I}$ and minimum distances for $(12,4)$ codes.}
	\label{table:example-design-comparison}
\end{table}

\section{Numerical Examples}
\label{sec:examples}
In this section we evaluate the performance of the proposed multi-kernel polar codes under the different designs. 
All the simulations determine the BLock Error Rate (BLER) of the codes for BI-AWGN channels under SCL decoding, usually for list size  $L=8$. 
To begin with, we evaluate the impact of the kernel order in the multi-kernel construction. 
Next, the impact of parameter $\psi$ of the hybrid design is studied. 
Then we compare multi-kernel polar codes to punctured and shortened polar codes of same length and dimension. 
Finally, we compare multi-kernel polar codes with standard codes of same length and dimension, namely with LDPC codes for 802.11n \cite{802_11n} and polar codes for 5G NR \cite{polar_5G}.

\begin{figure}[th]
\centering
	\includegraphics[width=0.48\textwidth]{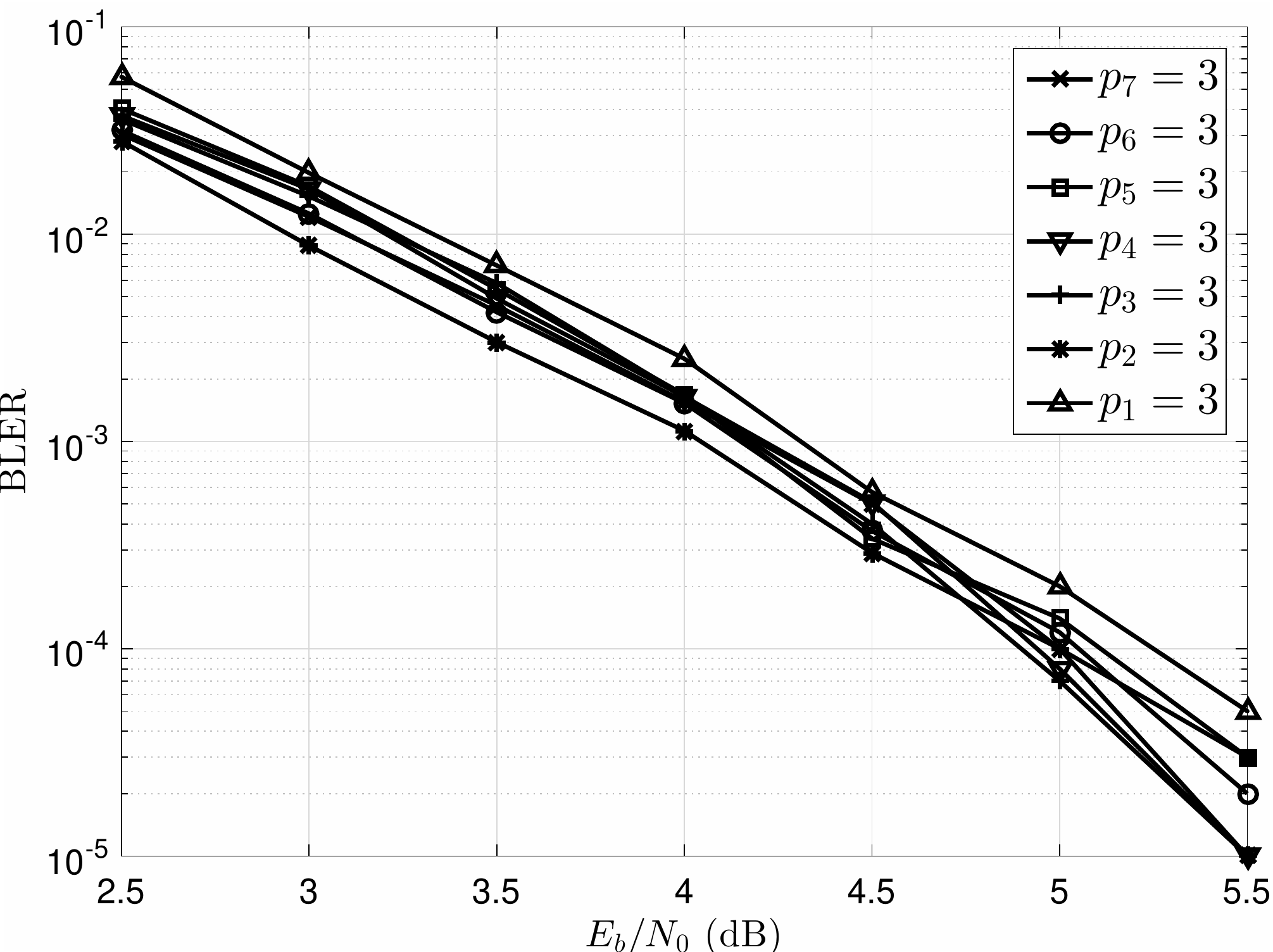}
	\caption{BLER performance of $(192,96)$ multi-kernel polar codes under reliability design for different position of the unique $T_3$ kernel and list size $L=8$.}
	\label{fig:kernel_ord}
\end{figure}

Figure~\ref{fig:kernel_ord} shows the BLER performance of multi-kernel polar codes of length $N = 192 = 2^6 \cdot 3$ and rate $R=1/2$, designed according to reliability, under SCL decoding with list size $L=8$. 
The transformation matrix of this code is constructed mixing 6 kernels $T_2$ and a single $T_3$ kernel. 
There are hence 7 possible configurations of \eqref{equ:GN-MK}, namely depending on the position of the $T_3$ kernel in the Kronecker product. 
According to Figure~\ref{fig:kernel_ord}, the performance gain between the best and the worst design for this code is about 0.5 dB. 
In particular, the best performance is obtained when $p_2=3$, i.e. when $T_3$ is placed in second position, while the worst performance is attained by switching the first two kernels of the best design, with $p_1=3$. 
Note also that the slope of the curves is different, which is due to differences in their distance properties. 
The metric proposed in Section~\ref{sec:reliability} for the selection of the kernel order suggests setting $p_3=3$, resulting in average performance at low SNR but quickly approaching the performance of $p_2=3$ at higher SNR. 
Simulations confirm that the kernel order has some impact on the performance of multi-kernel polar codes under reliability design; no systematic behavior could be identified at this stage.

\begin{figure}[th]
\centering
	\includegraphics[width=0.48\textwidth]{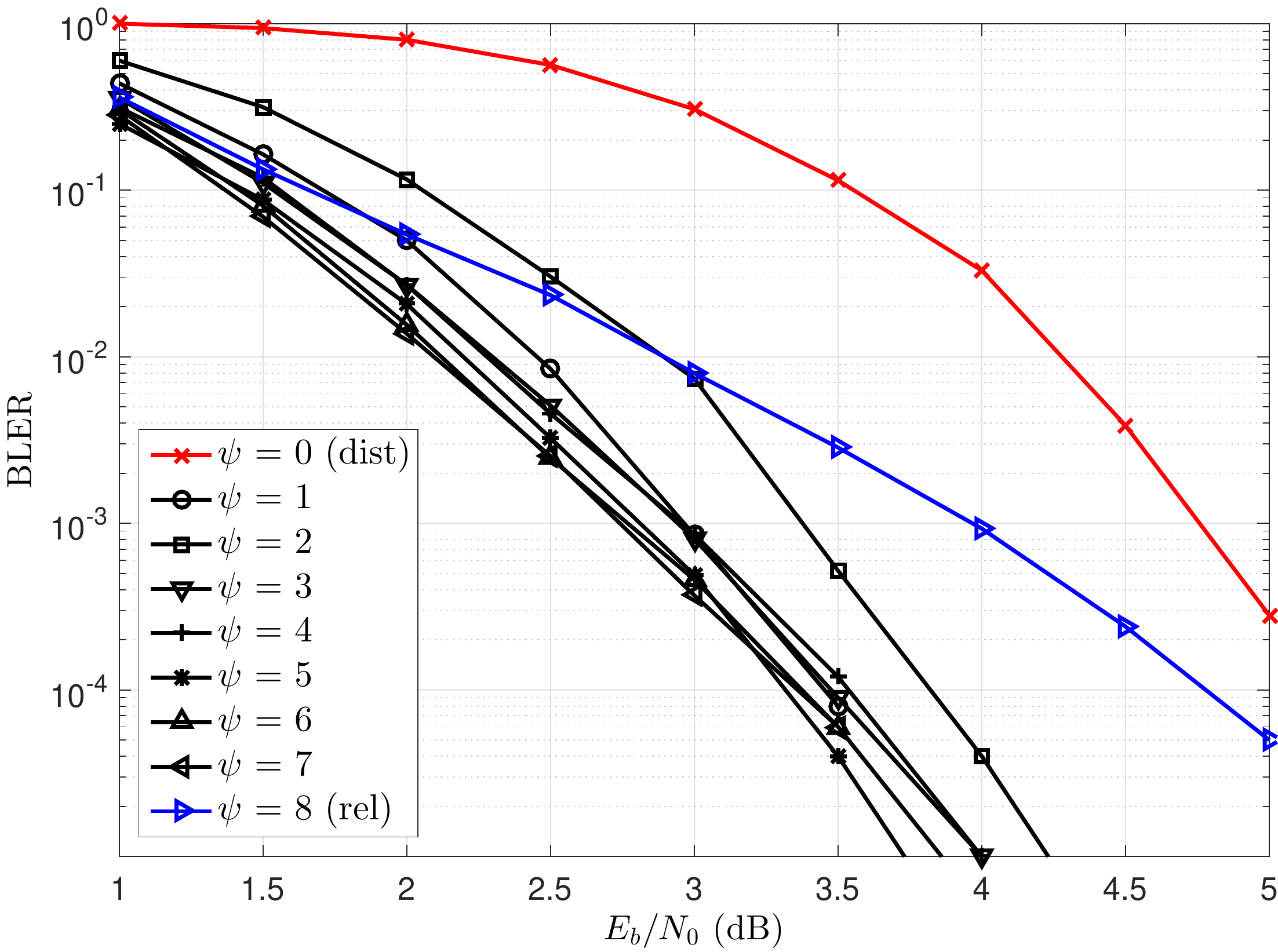}
	\caption{BLER performance of $(384,192)$ multi-kernel polar codes under hybrid design for different values of $\psi$ and $L=8$.}
	\label{fig:psi}
\end{figure}

Figure~\ref{fig:psi} shows the BLER performance of multi-kernel polar codes of length $N=384 = 2^7 \cdot 3$ and rate $R=1/2$ under hybrid design for different values of parameter $\psi$ introduced in Section~\ref{sec:hybrid}. 
All the simulations are performed under SCL decoding with list size $L=8$. 
In this case, the number of kernels composing the transformation matrix is $s=8$, and the single $T_3$ kernel is placed in the last position, having $p_8=3$. 
Parameter $\psi$ can hence span from $0$ to $8$, where the extreme cases $\psi=0$ and $\psi=8$ corresponding to distance and reliability designs, respectively, are highlighted with different colors. 
As expected for mid-length codes, distance design performs worse than reliability design, however the code is not long enough to have strong polarization. 
In this case, the mixed design offers an advantage over the other two designs due to higher flexibility. 
Simulations show that the performance strongly depends on the choice of parameter $\psi$; a clear pattern is not recognizable and needs further studies. 
In the following, we set $\psi = \lceil \frac{s-1}{2} \rceil$ as a rule of thumb for the hybrid design. 

\begin{figure*}
\begin{center}
\subfloat[$N = 144$]{\includegraphics[width=0.32\textwidth]{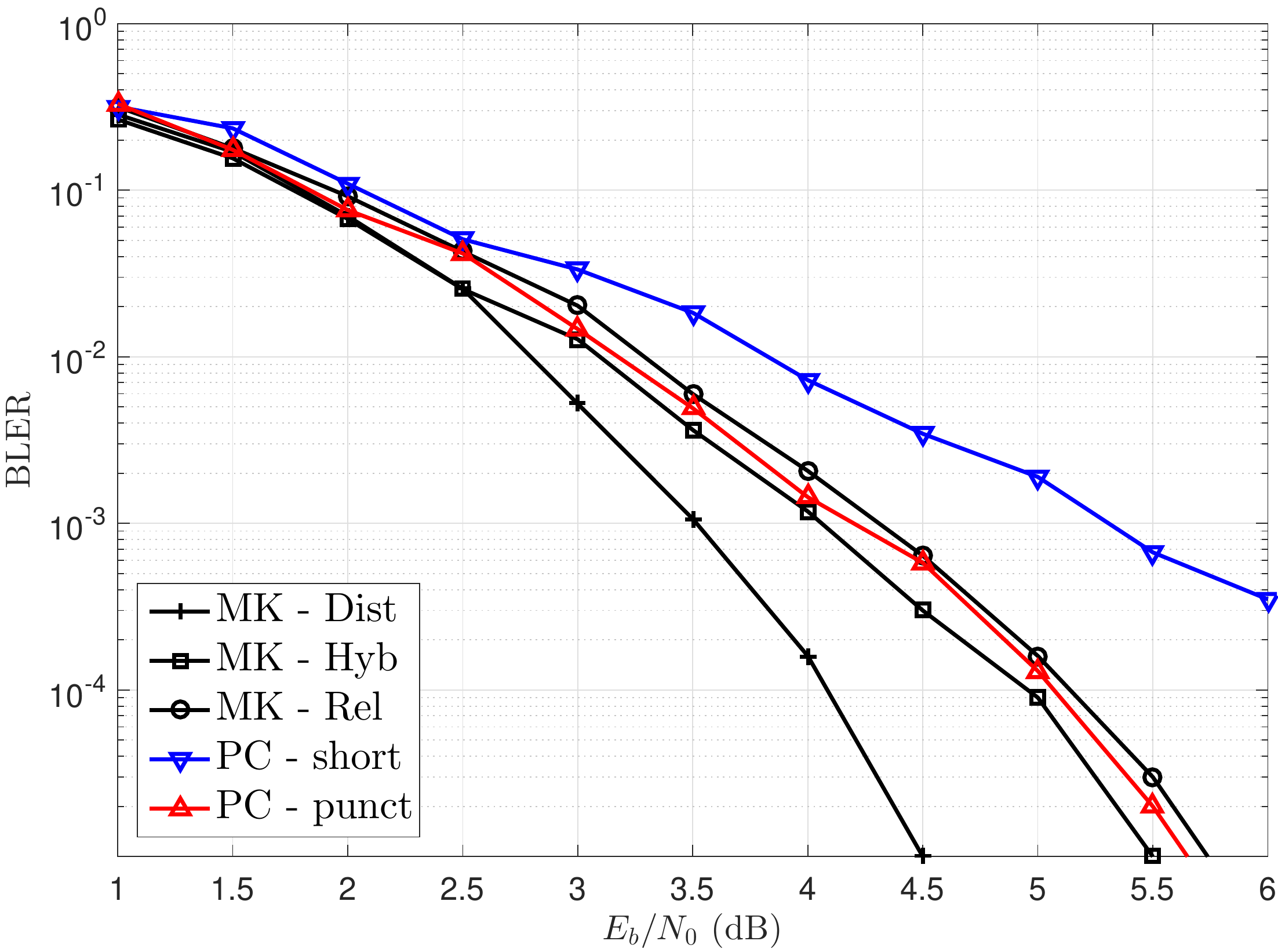} \label{fig:144_72}}
\subfloat[$N = 200$]{\includegraphics[width=0.32\textwidth]{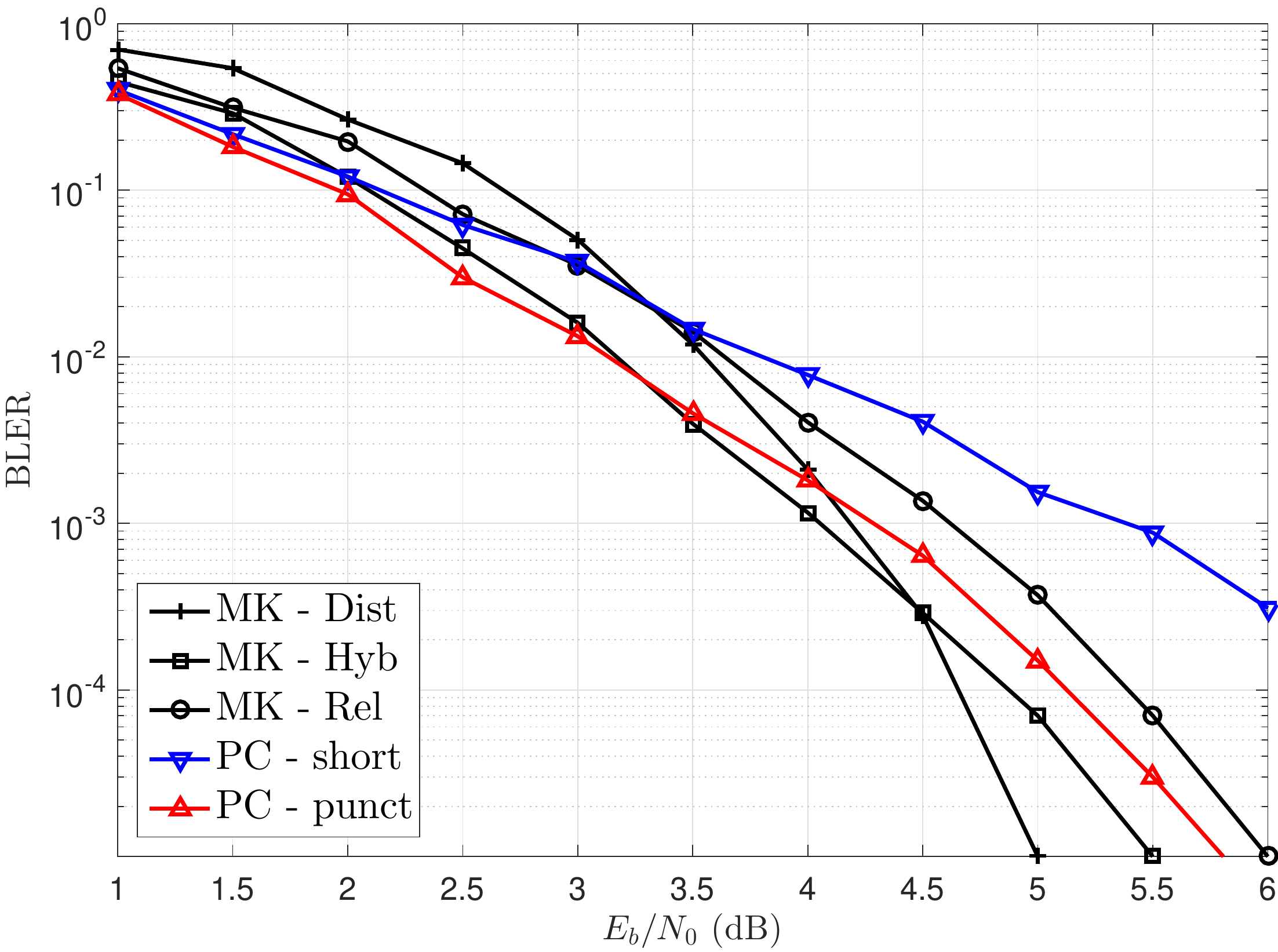} \label{fig:200_100}} 
\subfloat[$N = 90$]{\includegraphics[width=0.32\textwidth]{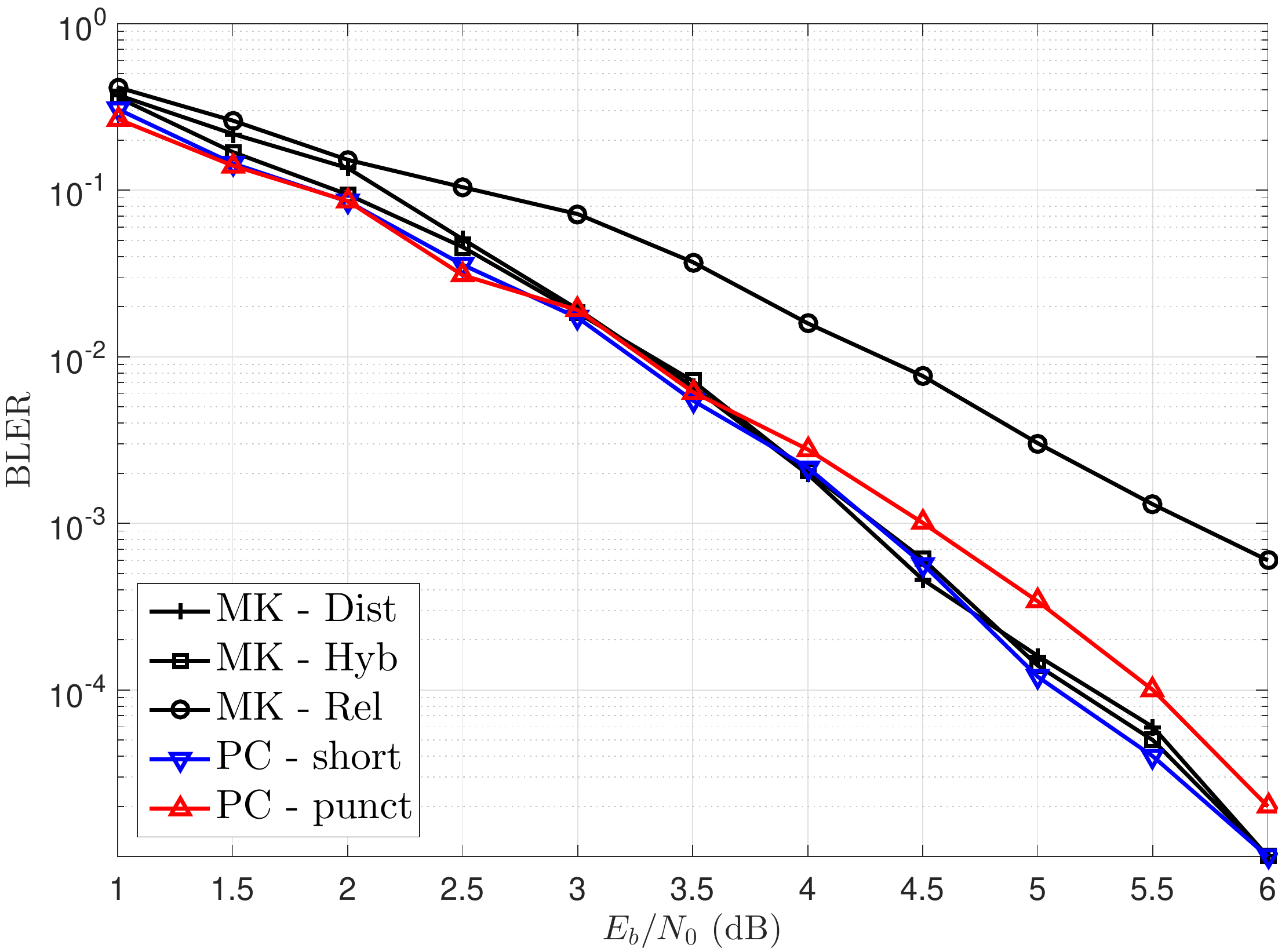} \label{fig:90_45}} 
\caption{BLER performance of multi-kernel polar codes compared with punctured \cite{chen_kai_punc} and shortened \cite{wang_liu} polar codes of rate $R = 1/2$ under SCL decoding with $L=8$.}
\label{fig:MK_vs_short}
\end{center}
\end{figure*}

Figure~\ref{fig:MK_vs_short} shows a comparison between the presented multi-kernel polar codes under various designs and rate-matched polar codes for rate $R=1/2$ under SCL decoding with $L=8$. 
Polar codes of length $N$ are generated from a mother polar code of length $M = 2^{\log2(\lceil N \rceil)}$ through puncturing according to \cite{chen_kai_punc} and shortening according to \cite{wang_liu}. 
The information sets of mother polar codes are hence calculated by either puncturing the first $M-N$ bits or shortening the last $M-N$ bits and calculating bit reliabilities through DE/GA \cite{punct_paper}. 
Code lengths are selected to have a wide range of possibilities. 

In Figure~\ref{fig:144_72} the code length is set to $N = 144 = 3^2 \cdot 2^4$. This length can be reached by multi-kernel polar codes with two $T_3$ kernels and four $T_2$ kernels, while it demands strong puncturing/shortening by $144$ bits to apply rate-matching from a mother polar code of length $M = 256$. 
Figure~\ref{fig:200_100} shows the performance of codes of length $N = 200 = 5^2 \cdot 2^3$, reached by multi-kernel polar codes using two $T_5$ kernels in conjunction to $T_2$ kernels. 
The last Figure~\ref{fig:90_45} mixes all the presented kernels showing the performance of codes of length $N = 90 = 5 \cdot 3^2 \cdot 2$. 

Overall, we can see that reliability design of multi-kernel polar codes behaves similarly to the best rate-matching strategy among puncturing and shortening. 
The sub-optimality of $T_5$ LLR equation $f^5_2$ has an impact on the performance of the reliability design for short codes, as shown in Figure~\ref{fig:200_100}. 
However, other designs are not impacted excessively by this issue, exhibiting good performance in this case. 
Moreover, distance design outperforms other designs for short codes, where minimum distance have still more impact than polarization effect; hybrid design permits a tradeoff between polarization and distance, always outperforming rate-matched polar codes. 

\begin{figure}[bth]
\centering
	\includegraphics[width=0.48\textwidth]{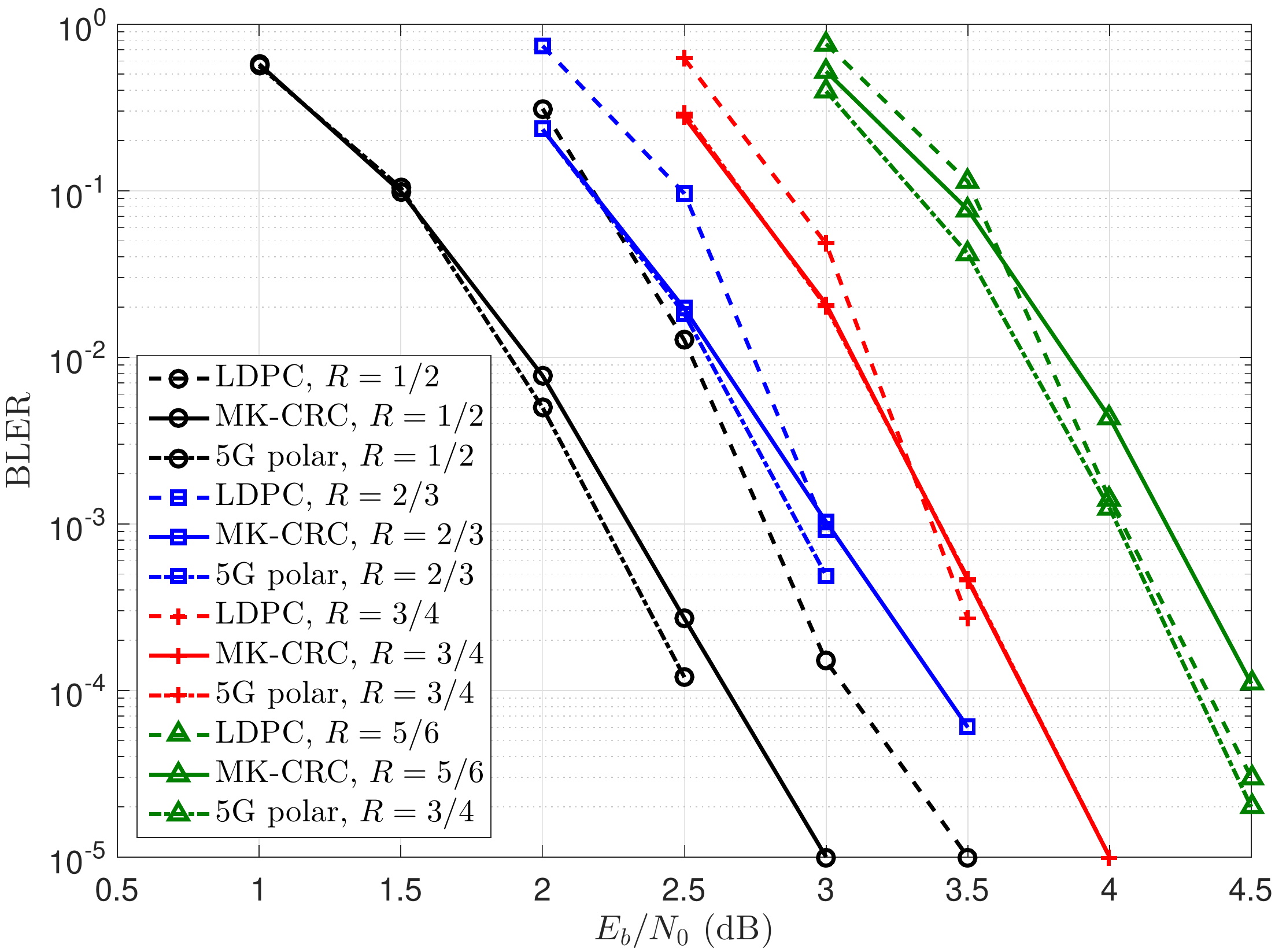}
	\caption{BLER performance comparison among multi-kernel polar codes with 8 CRC bits, LDPC codes in \cite{802_11n} and 5G polar codes \cite{polar_5G} for length $N = 1944$.}
	\label{fig:MK_vs_LDPC}
\end{figure}

Figure~\ref{fig:MK_vs_LDPC} shows a comparison among multi-kernel polar codes, LDPC codes of the 802.11n standard \cite{802_11n} and polar codes standardized in 5G \cite{polar_5G}. 
The 802.11n  standard specifies the three code lengths 1944, 1296 and 648, and the four code rates $1/2$, $2/3$, $3/4$ and $5/6$; we show simulation results for $N=1944$ and all four admissible rates. 
Multi-kernel polar codes are designed according to reliability, with the addition of 10 CRC bits to help SCL decoding \cite{list_decoding}. 
List size is set to $L=8$ for both multi-kernel polar codes and 5G polar codes; LDPC codes are decoded using a 10-iterations offset min-sum decoder. 
Results show that the proposed multi-kernel polar codes are comparable to state-of-the-art channel codes.

\section{Conclusions}
\label{sec:conclusions}
In this paper, we proposed a generalized polar code construction based on multiple kernels, termed multi-kernel polar codes.  
Though encoding and decoding resemble those of the original polar codes, as proposed by Arikan, multi-kernel polar codes provide various new design options.  
We presented such new code design principles based on reliability, distance, and a mix of those two as design criteria, coined as hybrid design, allowing to adapt the design to a given list length of the SCL decoder.  
The error-rate performance of multi-kernel polar codes was evaluated by simulations, resulting to be superior to state-of-the-art polar-code constructions, using puncturing or shortening methods, and state-of-the-art LDPC codes.

This paper focused on the information set design of multi-kernel polar codes.  
the design of optimal kernels or the optimization of hybrid design are not addressed in this paper and left for future research; the presented tools for analysis and design are believed to be useful for these purposes.

\bibliographystyle{IEEEbib}
\bibliography{polar_codes_bib}
\end{document}